\documentclass[eqsecnum,floats,aps,amsmath,amssymb,nofootinbib,showpacs]%
{revtex4}

\usepackage{amssymb,amsfonts,amsmath,amsopn,amsthm}
\usepackage{graphicx}
\usepackage{enumerate}
\usepackage{calc}

\usepackage{IHsymbols}

\begin{document}


\begin{flushright}
  IGPG-06/5-1
\end{flushright}

\preprint{\vbox{\baselineskip=12pt \rightline{IGPG-06/5-1}}}

\title{\bf Symmetric non-expanding horizons}

\author{Jerzy Lewandowski${}^{1,2,3}$}
  \email{lewand@fuw.edu.pl}
\author{Tomasz Pawlowski${}^{2,1}$}
  \email{pawlowsk@gravity.psu.edu}
\affiliation{
  ${}^{1}$Instytut Fizyki Teoretycznej, Uniwersytet Warszawski,\\ 
    ul. Ho\.{z}a 69, 00-681 Warsaw, Poland\\
  ${}^{2}$Institute for Gravitational Physics and Geometry\\
    Physics Department, Penn State, University Park, PA 16802,
    U.S.A.\\
  ${}^{3}$Perimeter Institute for Theoretical Physics, \\
    31 Caroline Street North, Waterloo, Ontario N2L 2Y5, Canada
}

\begin{abstract}
  Symmetric non-expanding horizons are studied in arbitrary
  dimension. The global properties -as the zeros  of infinitesimal 
  symmetries- are analyzed particularly carefully. For the class of
  NEH geometries admitting helical symmetry a quasi-local analog
  of Hawking's rigidity theorem is formulated and proved: the presence
  of helical symmetry implies the presence of two symmetries:  null,
  and  cyclic. 

  The results valid for arbitrary-dimensional horizons are next
  applied in a complete classification of symmetric NEHs in
  $4$-dimensional space-times (the existence of a $2$-sphere
  crossection is assumed). That classification divides possible NEH
  geometries into classes labeled by two numbers - the dimensions of,
  respectively, the group of isometries induced in the horizon base
  space and the group of null symmetries of the horizon.
\end{abstract}


\maketitle


\section{Introduction}

A non-expanding horizon (NEH) is a null, non-expanding $n-1$-surface
contained in an $n$ dimensional spacetime of signature
$(-,+,...,+)$. What distinguishes a NEH among other non-expanding
null surfaces is its topology,  assumed to be  the Cartesian product
of a compact spacelike crossection $\tilde{\ih}$ with a null
interval $\cal I$.  The theory of NEH in $n=4$ dimensions was proposed
by Ashtekar et. al. \cite{abf,abl-m,abl-g,abdfklw} as a quasi local
generalization of the black hole theory. The framework and many
results were generalized to  an arbitrary spacetime dimension $n>2$
(\cite{adw} the $n=3$ case, and \cite{lp-g,klp}  the $n\ge 3$ case).
The non-rotating (see below) NEH horizons were defined by Newman and
Pejerski \cite{newman}. Null and compact surfaces considered
cosmological horizons were studied by \cite{ismo}.

A short outline of the published results should be started with a
remark, that the (even local) existence of non-stationary vacuum
spacetimes admitting NEHs \cite{l,c} came as surprise to several
experts in the black hole theory. The theory of NEH can be divided
into two chapters: Geometry and Mechanics. The scope of this paper
is Geometry, hence for  Mechanics we refer the reader to
\cite{abl-m,ak-rev,klp}.

The geometry of a NEH $\ih$ in spacetime which satisfies the
Einstein equations (with or without a cosmological constant) and the
weak energy condition, consists of the induced: (degenerate) metric
tensor $q$ on $\ih$ , and the covariant derivative $D$ in the bundle
tangent to $\ih$.\footnote{That unique connection is a peculiar
property of the
  non-expanding and shear-free null surfaces. It is not shared by
  generic null surfaces.}
The Einstein equations impose  constraints on the geometry. The
constraints are explicitly soluble. The structure of a general
solution was studied in \cite{abl-g,lp-g}. For every solution,  due
to the geometric generalization of ``the zeroth law of black hole
thermodynamics'', there is the invariantly defined  rotation 2-form.
Other invariants \cite{abl-g}  can be used to construct invariant
coordinates in a spacetime neighborhood of a given generic $\ih$.
Yet another invariant,  one extensively used in this paper, is the
Jezierski-Kijowski vector field \cite{JK}. This vector field is null
and defined by the NEH geometry uniquely up to rescalings by a
constant factor.

We call a NEH $\ih$ symmetric, if there exists a vector field $X$
defined on $\ih$, such that its local flow is a local symmetry of
the NEH geometry $(q,D)$. The vector field itself is called an
infinitesimal symmetry. If the infinitesimal symmetry $X$ is null
but nowhere vanishing, we say that it defines on $\ih$ an isolated
horizon (IH) structure, or just briefly we call $\ih$ an isolated
horizon.

All the  non-extremal (that is, in this case, such that
$D_XX\not=0$) IH geometries, solutions of the vacuum Einstein
constraints were constructed explicitly \cite{abl-g,lp-g}. A free
data can be defined on any space-like crossection $\slc$ of $\ih$.
It consists of a metric tensor $\tilde{q}$ and some differential
1-form $\tilde{\omega}$, a potential for the rotation 2-form.
Compared with the parametrization of the Kerr metric family,
$\tilde{q}$ is a generalization of the radius of the Kerr black hole
horizon, and  $\tilde{\omega}$ is the generalization of the angular
momentum.

The question of what local properties distinguish the Kerr NEH
geometry was raised in \cite{lp-kerr} and solved in the following
way. In the $n=4$ dimensions, the conditions that at an IH the
spacetime Weyl tensor be of the Petrov type D, whereas the Ricci
tensor vanish, are equivalent to the vanishing of certain invariant
of the IH geometry.  The only axial solutions to that condition are
the geometries defined by the family of the Kerr spacetimes. That
result provides a geometric, coordinate invariant, local
characterization of the NEH whose geometry coincides with that of
the Kerr black hole.

In the case of an  extremal (i.e. non non-extremal) IH, the Einstein
constraints take the form of a non-linear equation imposed  on the
pair $(\tilde{q},\tilde{\omega})$, the projection onto $\slc$ of the
metric $q$ and the rotation potential $\omega$ respectively
\cite{abl-g,lp-g}. In the vacuum case, the extremal IH equation
reads
\begin{equation}\label{eq:in-extr}
  \tD_{(A}\twI_{B)}\ +\ \twI_{(A} \twI_{B)}\
  -\ \frac{1}{2}\tilde{\Ric}_{AB}\ =\ 0,
\end{equation}
where $\tD_A$ and $\tilde{\Ric}_{AB}$ are the torsion free covariant
derivative and the Ricci tensor, respectively, of the metric
$\tilde{q}$. In $n=4$ dimensions, the equation has solutions only if
the topology of a spacelike section of the IH is either that of
2-torus or 2-sphere \cite{plj}. In the first case, the only solution
is the trivial one. In the second case, the following two results
are known. According to the first one  \cite{ex}, the only axial
solutions are those defined by the extremal Kerr spacetimes. The
second result due to Chru\'sciel, Real and Tod \cite{crt} is that
there are no non-rotating  solutions. A short proof of this result
is also hidden in the NEH literature \cite{ex} and \cite{abl-g}
(however, the authors failed to notice that conclusion) and we will
demonstrate it at the end of subsection \ref{sec:nullsym-ex-evac}.

In the $n=4$ case, an {\it a priori} unexpected relation between the
extremal IH equation \eqref{eq:in-extr} on the one hand, and the
Kundt constraint (31.15ab, 31.16ab) in \cite{exact} on the other
hand, was found \cite{plj}. Via the relation, every solution of the
extremal IH equation can be used to construct a vacuum spacetime, an
exact solution of the Einstein equations which belongs to the
Kundt's class. In particular, the spacetime was constructed whose
topology is $S^2\times\re\times\re$ and every surface  $S^2\times
\re\times\{r\}$ is a Killing horizon ($S^2$ is a $2$-sphere and $r$
ranges $\re$.)

In the early stages of developing the NEH theory a lot of attention
was payed to the issue of the uniqueness of  infinitesimal null
symmetry. The hope was, that given a NEH, if a null infinitesimal
symmetry exists, it should be  unique  modulo re-scalings by a
constant factor. The result of the research on that issue was the
discovery  of NEH admitting 2-dimensional group of null symmetries
\cite{abl-g,lp-g}. Explicit examples were constructed out of the
extremal Kerr horizon \cite{ex}.

The goal of this paper is a systematic analysis
of the symmetric NEHs.  All our considerations are global in the
sense of the manifold $\ih$.

The basic definitions and geometric properties of the NEHs used in
this paper are recalled in section \ref{sec:intro-geom}.

A general result of section \ref{sec:symm-basic} (see proposition
\ref{prop:symm-ext}) is that every symmetric NEH is a segment of an
(abstract, not necessarily embedded) symmetric NEH whose null curves
are complete in any affine parametrization. Moreover, on that maximal
analytic extension of a given symmetric NEH, the infinitesimal
symmetry generates a group of globally defined symmetry maps.
Therefore, in the main part of the paper, starting from
Section \ref{sec:symm-gen-null} through out the whole paper {\it we
  identify each NEH $\ih$  with its maximal analytic extension.}
The fact that in general the extension is not embedded in the
space-time should not lead to any confusion.

The null symmetries of a symmetric NEH considered in the previous
works, were assumed to act non-trivially on any null curve. In the
current work we relax that assumption and study the zeros of all the
possible null infinitesimal symmetries.  The new results on the null
symmetries are combined with the previous ones  \cite{abl-g,lp-g,ex}

The most interesting new result is a generalization of the Hawking
rigidity theorem to the NEH context. We prove, that every helical
NEH necessarily is cyclic (or even axial), and admits a null
infinitesimal symmetry. Our generalization extends in two
directions: $(i)$ from globally defined black hole to  quasi locally
defined NEH, and $(ii)$  from $n=4$ to arbitrary $n>2$. In the
literature, Hawking's rigidity theorem was also generalized to
compact, null surfaces in \cite{ismo}.

The two results enlisted above lead us 
to a complete classification (discussed in subsection
\ref{sec:4D-class}) of the symmetric NEHs in the $n=4$ dimensional
spacetime and the spherical topology of a space-like cross-section 
case.

In this introduction we kept track of the works on the NEH geometry
closely related to our current work. However, there is also the
interesting literature ranging from papers discussing various
mechanical approaches to the NEHs \cite{booth}, to the works dealing
with similar study of other surfaces, whose scopes occasionally
overlap with ours \cite{cjk,hayward,ak-dh,f-dh,nr}.

\section{Geometry of a non-expanding horizon}
  \label{sec:intro-geom}

\subsection{Non-expanding null surfaces}
  \label{sec:nes-def}
In this section we introduce the notation, recall the definition and
properties of non-expanding horizons \cite{abdfklw,abl-g}. The related
calculations concerning the general $n$-dimensional case can be found
in \cite{lp-g}.

\subsubsection{Definition, the induced metric}

Consider an $(n-1)$-dimensional null surface $\ih$ embedded in an
$n$-dimensional spacetime $\M$. The spacetime metric tensor
$g_{\mu\nu}$ of the signature $(-,+,\cdots,+)$ is assumed to satisfy
the Einstein field equations (possibly with matter and cosmological
constant). We will denote the degenerate metric tensor induced at
$\ih$ by $q_{ab}$. The subbundle of the tangent bundle $T(\ih)$
defined by the null vectors will be denoted by $L$ and referred
to as the null direction bundle. Given a vector bundle $P$, the set
of sections will be denoted by $\Gamma(P)$.

\begin{defn}\label{def:nes}
  Given a null surface $\ih$ embedded in spacetime satisfying the
  Einstein field equations it is called a {\it non-expanding null
  surface} (NES) if for every point $x\in\ih$ the
  expansion of some nontrivial null vector $\ell^a$
  tangent to $\ih$ at $x$ vanishes.
\end{defn}

The Raychaudhuri equation implies that provided the energy-momentum
tensor of matter fields satisfies at $\ih$ the following energy
condition
\begin{equation}\label{eq:Tll>0}
  T_{ab}\ell^a\ell^b\ \geq\ 0 \ ,
\end{equation}
(with $T_{ab}$ being the pull-back of the spacetime energy-momentum
$T_{\mu\nu}$ onto $\ih$) the flow $[\ell]$ preserves the degenerate
metric $q$
\begin{equation}\label{eq:lie_q}
 \lie_{\ell}q_{ab}\ =\ 0 \ ,
\end{equation}
and the component $\Ricn{n}_{\ell\ell}$ of the spacetime Ricci tensor
vanishes. The condition \eqref{eq:Tll>0} will be further referred to
as the {\it Weaker Energy Condition}.

The property \eqref{eq:lie_q} above combined with  $\ell^aq_{ab}=0$
means that, locally $q_{ab}$ is the pullback of a certain metric
tensor field $\hq_{AB}$ defined on an $(n-2)$-dimensional manifold
$\bas'$. The manifold $\bas'$ is the space of the null curves tangent
to $\ih$ contained in a given (sufficiently small) neighborhood
$\ih'\subset\ih$ open in $\ih$, and the map is the natural projection,
\begin{subequations}\label{eq:bas}\begin{align}
  \spi:\ih'\ &\rightarrow\ \bas' \ , &
  q_{ab}\ &=\ \spi^*\hq_{AB} \ . \tag{\ref{eq:bas}}
\end{align}\end{subequations}

\subsubsection{The covariant derivative}

If at a given NES $\ih$ the matter fields satisfy the Weaker Energy
Condition \eqref{eq:Tll>0} then for any vector fields $X,Y$, sections
of the tangent bundle $T(\ih)$, the covariant derivative $\nabla_XY$
is again a vector field tangent to $\ih$. Therefore, there is an
induced connection $D_a$ in $T(\ih)$, such that for every pair of
vector fields $X,Y\in \Gamma(T(\ih))$
\begin{equation}
  D_XY^a\ :=\ \nabla_XY^a \ .
\end{equation}
For a covector $W_a$, a section of the dual bundle $T^*(\ih)$,
the derivative $D_XW_a$ is determined by the Leibnitz rule,
\begin{equation}
  Y^aD_XW_a\ =\ D_X(Y^aW_a) - (D_XY^a)W_a \ .
\end{equation}
Obviously, the derivative $D_a$ is torsion free and annihilates
the degenerate metric tensor $q_{ab}$,
\begin{subequations}\label{eq:D-tfree}\begin{align}
  D_aD_b f\ &=\ D_bD_a f \ ,  &
  D_aq_{bc}\ &=\ 0 \ ,  \tag{\ref{eq:D-tfree}}
\end{align}\end{subequations}
for every function $f$.

\subsubsection{The rotation $1$-form}

The covariant derivative $D_a$  induced on $\ih$  preserves the null
direction bundle $L$. It implies that the derivative $D_a\ell^b$ is
proportional to $\ell^b$ itself,
\begin{equation}\label{eq:omega_def}
  D_a\ell^b\ =\  \w{\ell}{}_a\ell^b \ ,
\end{equation}
where  $\w{\ell}{}_a$ is a 1-form defined uniquely on this subset
of $\ih$ on which $\ell\neq0$ is defined. We call $\w{\ell}{}_a$
the rotation 1-form potential (see \cite{abl-g,lp-g}).

The evolution of $\w{\ell}{}_a$ along the surface $\ih$ upon the null
flow is responsible for the $0$th Law of the non-expanding horizon
thermodynamics:
\begin{equation}\label{eq:pre0th}
  \lie_{\ell}\w{\ell}{}_a\ =\ D_a\sgr{\ell} \ +\ \Ricn{n}_{ab}\ell^b
\end{equation}
where the {\it surface gravity} $\sgr{\ell}$ is given by $\w{\ell}_a$
as follows
\begin{equation}\label{eq:sgr}
  \sgr{\ell}\ =\ \w{\ell}{}_a\ell^a \ .
\end{equation}

We also  strengthen the energy conditions imposed on $T_{\mu\nu}$,
namely we most often assume in this paper that the following holds:
\begin{cond}\label{c:energy}(Stronger Energy Condition)
  At every point of the surface $\ih$, for every future oriented null
  vector $\ell$ tangent to $\ih$, the vector
  \begin{equation}
    -T^{\mu}{}_{\nu}\ell^{\nu}
  \end{equation}
  is causal, that is
  \begin{equation}\label{eq:T_cau}
    g^{\mu\nu}T_{\mu\alpha}\ell^{\alpha}T_{\nu\beta}\ell^{\beta}\
    \leq\ 0 \ ,
  \end{equation}
  and future oriented.
\end{cond}
This condition implies automatically the previous one
$T_{\ell\ell}\geq 0$. Also (via the Einstein field equations) it
imposes the vanishing of certain Ricci tensor components at
$\ih$, namely
\begin{equation}
  \Ricn{n}_{ab}\ell^b\ =\ 0 \ .
\end{equation}
The evolution of the rotation potential (given by
\eqref{eq:pre0th}) is then described by the following theorem:
\begin{theorem}[The $0$th Law]\label{thm:0th}
  Suppose $\ih$ is an $(n-1)$-dimensional, non-expanding, null
  surface; suppose that the Einstein field equations hold on $\ih$
  with a cosmological constant and with the matter fields which
  satisfy the Stronger Energy Condition \ref{c:energy}.
  Then, for every null vector field $\ell^a$ defined on and tangent to
  $\ih$, the corresponding rotation 1-form potential $\w{\ell}{}$ and
  the surface gravity $\sgr{\ell}$ satisfy the following constraint:
  \begin{equation}\label{eq:0th}
    \lie_{\ell}\w{\ell}{}_a\ =\ D_a\sgr{\ell} \ .
  \end{equation}
\end{theorem}

Theorem \ref{thm:0th} tells us, that there is always a choice of
the section $\ell$ of the null direction bundle $L$ such that
$\w{\ell}{}$ is Lie dragged by $\ell$. For,  we can always find a
non-trivial section $\ell$ of $L$ such that $\sgr{\ell}$ is
constant. The relation with the original $0$th Law of black hole
thermodynamic goes the other way around. Indeed, if the vector
field $\ell^a$ admits an extension to a Killing vector defined in
a neighborhood of $\ih$, then $\w{\ell}{}$ is Lie dragged by the
flow, therefore the left hand side is zero, hence $\sgr{\ell}$ is
necessarily (locally) constant.

Upon rescalings $\ell\mapsto\ell'=f\ell$ (where $f$ is a real function
defined at $\ih$) of the section $\ell^a$ of $L$ the rotation $1$-form
changes as follows
\begin{equation}\label{eq:omega'}
  \w{\ell'}{}_a\ =\ \w{\ell}{}_a + D_a \ln f \ .
\end{equation}
Therefore its exterior derivative (in the sense of the
manifold $\ih$) called {\it the rotation $2$-form} is independent
of the choice of a null vector field $\ell\in\Gamma(L)$, i.e.
\begin{equation}\label{eq:Omega}
  \Omega_{ab}\ :=\ D_a\w{\ell}{}_b - D_b\w{\ell}{}_a\
    =\ D_a\w{\ell'}{}_b - D_b\w{\ell'}{}_a \ .
\end{equation}

\subsection{Geometry of a NES and the constraints}
  \label{sec:nes-geom}

Given a non-expanding null surface $\ih$, the pair $(q_{ab}, D_a)$,
that is the induced degenerate metric and, respectively, the induced
covariant derivative are referred to as the geometry of $\ih$. By a
`constraint' on the non-expanding surface geometry we mean here every
geometric identity $\mathcal{F}(q_{ab}, D_a,\Ricn{n}_{\alpha\beta})=0$
involving the geometry $(q_{ab},D_a)$ and the spacetime Ricci tensor
at $\ih$ only. Part of the constraints is already solved by the
conclusion that $q_{ab}$ be Lie dragged by every null flow generated
by a null vector field $\ell$ tangent to $\ih$ (see
\eqref{eq:lie_q}). Another example of a constraint is the $0$th Law
(\ref{eq:pre0th}, \ref{eq:0th}). A complete\footnote{Among all the
  components of the Einstein tensor only its pullback to $\ih$ can be
  involved in a constraint. It will be shown further that its
  value is determined by the commutator $[\lie_\ell,D_a]$. The
  remaining components involve transversal derivatives of the
  components of $\nabla_{\mu}$ (where the number of determined
  transversal derivatives is equal to the number of the remaining
  components of the Einstein tensor).
} set of the functionally independent constraints is formed by
$\lie_\ell q_{ab}=0$ and by an identity satisfied by the commutator
$[\lie_\ell,D_a]$, where $\ell$ is a fixed, non-vanishing section of
the null direction bundle $L$.

We turn now to the second identity mentioned above. The commutator
itself is proportional to $\ell^b$
\begin{equation}\label{eq:[l,D]1}
  [\lie_\ell,D_a]X^b\ =\ \ell^b N_{ac}X^c \ ,
\end{equation}
where the tensor $N_{ab}$ can be expressed by the rotation potential,
its derivative and the spacetime Ricci tensor
\begin{equation}\label{eq:[l,D]2}
  N_{ac}\ =\  D_{(a}\w{\ell}{}_{c)}\ +\ \w{\ell}{}_a\w{\ell}{}_c
  + \frac{1}{2}\left(\Ricn{n}_{ac}-\spi^*\Ricn{n-2}_{ac}\right) \ .
\end{equation}
The contraction of (\ref{eq:[l,D]1}, \ref{eq:[l,D]2}) with $\ell^a$ is
equivalent to \eqref{eq:pre0th} whereas the meaning of the remaining
part of the constraint (\ref{eq:[l,D]1}, \ref{eq:[l,D]2}) is explained
in the next sub-subsection after we itemize the derivative $D_a$ into
components.

\subsubsection{Compatible coordinates, foliations}

Further description of the elements of the covariant derivative
$D_a$ induced on a null, non-expanding surface $\ih$, and its relation
with the spacetime Ricci tensor require an introduction of an extra
local structure on $\ih$.

Given a a nowhere vanishing local section $\ell^a$ of the null
direction bundle $L$ one can define in the domain of $\ell^a$ a real
function $v$ compatible with $\ell^a$, that is such that
\begin{equation}\label{eq:def_v}
  \ell^a D_a v\ =\ 1 \ .
\end{equation}
The function $v$ referred to  as {\it a coordinate compatible
with $\ell$} defines on $\ih$ a covector field
\begin{equation}
  n_a\ :=\ -D_av \
\end{equation}
which is:
\begin{enumerate}[ (i)]
  \litem \label{it:n_norm}
    normalized in the sense that
    \begin{equation}\label{eq:n_norm}
      \ell^a n_a\ =\ -1 \ ,
    \end{equation}
    and
  \litem \label{it:n_fol}
    is orthogonal to the constancy surfaces $\slc_v$ of the
    function $v$ (referred to as {\it slices}).
\end{enumerate}
The family of he slices  is preserved by the null flow of $\ell$, and
so is $n_a$,
\begin{equation}\label{eq:lie-n}
  \lie_\ell n_a\ =\ 0 \ .
\end{equation}

At every point $x\in \ih$,  the tensor
\begin{equation}\label{eq:g_dec}
  \tq^{a}{}_{b}\ :=\ \delta^{a}{}_{b} + \ell^{a}n_{b}
\end{equation}
defines the orthogonal to $\ell^a$ projection
\begin{equation}
  T_x(\ih)\ \ni\ X^a\
  \mapsto\  \tilde{X}^a\ =\ \tq^{a}{}_{b}X^b\ \in\ T_x(\slc_v) \ .
\end{equation}
onto the tangent space $T_x(\slc_v)$, where $\slc_v$ is the
slice passing through $x$.\footnote{Instead of $\tilde{X}^a$
  we will write $\tilde{X}^A$, according to the index notation
  explained in Introduction.}
Applied to the covectors, elements of $T^*_x\ih$, on the other hand,
$\tq^{a}{}_{b}$ maps each of them into the pullback onto
$\slc_v$,\footnote{The result will be also
  denoted by by using a capital Latin index, as for example
  $\tilde{Y}_A$.}
\begin{equation}\label{tilde}
  T_x^*\ih\ \ni\ Y_a\ \mapsto\
  \tilde{Y}_a\ :=\ \tq^{b}{}_{a}Y_b\ \in\ T^*_x\slc_v \ .
\end{equation}

The field $n_a$ could be  extended to a section of the pullback
$T_\ih^*\M$ to $\ih$ of the cotangent bundle $T^*\M$,  by the
requirement that
\begin{equation}\label{eq:n_null}
  g^{\mu\nu}n_{\mu}n_{\nu}\ =\ 0 \ .
\end{equation}
Hence $n_a$ can be thought of as a transversal to $\ih$ null
vector field from the spacetime point of view.

\subsubsection{The components of $D_a$}

Each slice $\slc_v$ of the foliation introduced above is equipped with
the induced metric tensor $\tq_{AB}$ defined by the pullback of
$q_{ab}$ (and of $g_{\alpha\beta}$) to $\slc_v$. Denote by $\tD_A$ the
torsion free and metric  covariant derivative determined on $\slc_v$
by the metric tensor $\tq_{AB}$. All the slices are naturally
isometric.

The covector field $n_a$ gives rise to the following symmetric tensor
defined on $\ih$,
\begin{equation}\label{eq:S_def}
  S_{ab}\ :=\ D_an_b \ .
\end{equation}

Given the  structure introduced  previously  on $\ih$ locally (the
null vector field $\ell^a$, the foliation by slices $\slc_v$ and the
covector field $n_a$), the  derivative $D_a$ defined on $\ih$  is
determined by the following information
\begin{enumerate}[ (i)]
  \litem the torsion free covariant derivative $\tD_A$ corresponding
    to the Levi-Civita connection of the induced metric tensor
    $\tq_{AB}$ ,
  \litem the rotation 1-form potential $\w{\ell}_a$, and
  \litem a symmetric tensor $\tS_{AB}$ defined in each slice $\slc_v$,
    by the pullback of $D_an_b$,
    \begin{equation}\label{eq:SAB}
      \tS_{AB} \ =\ \tq^a{}_A\tq^b{}_BS_{ab} \ , \\
    \end{equation}
    and referred to the transversal expansion-shear tensor.
\end{enumerate}
Due to the normalization \eqref{eq:n_norm} the contraction of the
tensor with the null normal to $\ih$ is equal to:
\begin{equation}\label{eq:Sl}
  \ell^a S_{ab}\ =\ \w{\ell}{}_b \ .
\end{equation}

The constraint (\ref{eq:[l,D]1},\ref{eq:[l,D]2}) for $D_a$ can be now
expressed via the Lie derivative of $S_{ab}$ through the equality
\begin{equation}\label{eq:com_form}
  N_{ab}\ =\ \lie_{\ell}S_{ab} \ ,
\end{equation}
giving the following evolution equation for $S_{ab}$
\begin{equation}\label{S_evo_full}
  \lie_{\ell}S_{ab}\ =\ D_{(a}\w{\ell}_{b)} + \w{\ell}_a\w{\ell}_b
                    - \fracs{1}{2}\Riem{n}_{c(ab)}{}^d\ell^c n_d \ .
\end{equation}
The contraction of the above expression with $\ell^a$ reproduces the
$0$th Law, whereas the remaining component (the pullback of
$\lie_{\ell}S_{ab}$ onto a slice $\slc_v$) determines the evolution of
the transversal expansion-shear tensor $\tS_{AB}$,
\begin{equation}\label{eq:S_ev}
  \lie_{\ell}\tS_{AB}\ =\ -\sgr{\ell}\tS_{AB} + \tD_{(A}\tw{\ell}_{B)}
    + \tw{\ell}_A\tw{\ell}_B - \fracs{1}{2}\,\Ricn{n-2}{}_{AB}
    + \fracs{1}{2}\tRicn{n}{}_{AB} \ ,
\end{equation}
where tilde consequently means the projection (\ref{tilde}), and
$\Ricn{n-2}{}_{AB}$ is the Ricci tensor of the metric tensor
induced in  slice $\slc_v$  (since locally, every slice $\slc_v$ is
naturally isometric with the space of the null curves $\hat{\ih}'$
equipped with the metric tensor $\hat{q}_{AB}$ we denote the
corresponding Ricci tensors in the same way).

\subsection{Non-expanding horizons}
  \label{sec:neh}

\begin{defn}\label{def:neh}
  A non-expanding null surface  $\ih$ in an $n$ dimensional
  spacetime $\M$ is called a non-expanding horizon (NEH) if there is
  an embedding
  \begin{equation}
    \bas''\times\I\ \rightarrow\ \M
  \end{equation}
  such that:
  \begin{enumerate}[ (i)]
    \litem $\ih$ is the image,
    \litem $\bas''$ is an $n-2$ dimensional compact and
      connected\footnote{In the case $\bas''$ is not connected all the
        otherwise global constants (like surface gravity) remain
        constant only at maximal connected components of the
        horizon.} manifold,
    \litem $\I$ is the real line,
    \litem for every maximal null curve in $\ih$ there is
      $\hat{x}\in\bas''$ such that the curve is the image of
      $\{\hat{x}\}\times \I$.
  \end{enumerate}
\end{defn}
The {\it base space} $\bas$ defined as the space of all the maximal
null curves in $\ih$ can be identified with the manifold $\bas''$
given an embedding used in definition \ref{def:neh}. That embedding is
not unique, however the manifold structure defined in this way on
$\bas$ is unique. There is also a uniquely defined projection
\begin{equation}\label{eq:NEH-pi-def}
  \spi: \ih\ \rightarrow\ \bas \ ,
\end{equation}
onto the horizon base space.

As non-expanding horizons are just a special class of non-expanding
null surfaces, all the properties and structures developed for NESs
in subsections \ref{sec:nes-def}, \ref{sec:nes-geom} apply in to
NEHs. In particular (as the space $\bas'$ is now exactly the horizon
base space) $\bas$ is equipped with a metric tensor $\hq_{AB}$ such
that
\begin{equation}\label{eq:hq}
  q_{ab}\ =\ (\spi^*\hq)_{ab} \ ,
\end{equation}
with $\spi$ being the projection defined via \eqref{eq:NEH-pi-def}.
The tensor $\hq_{AB}$ will be referred to at the {\it projective
  metric}.

Through out of the remaining part of the article we will restrict our
considerations to non-expanding horizons only. Given a NEH $\ih$ there
exists a globally defined, nowhere vanishing null vector field $\ell^a$
tangent to it. In particular, there is a vector field $\ell_o^a$ of
the identically vanishing surface gravity, $\sgr{\ell_o}=0$. There
is also a null vector field $\ell^a$ of $\sgr{\ell}$ being an
arbitrary constant,\footnote{ The first one, $\ell_o$  can be defined
  by fixing appropriately affine parameter $v$ at each null curve in
  $\ih$. Then, the second vector field is just $\ell=v\ell_o$.}
\begin{equation}\label{eq:kappaconst}
  \sgr{\ell}\ =\ \const \ .
\end{equation}
The vector field $\ell^a$ can vanish in a harmless (for our
purposes) way on an $(n-2)$-dimensional section of $\ih$ only.

In the remaining part of this subsection, $\ell^a$ ($\ell_o{}^a$)
denotes a null vector field defined on and tangent to $\ih$, such
that \eqref{eq:kappaconst} (such that $\sgr{\ell_o}=0$). We will also
use a coordinate $v$ compatible with the vector field $\ell^a$ (
$\ell^aD_av=1$ ), and the covector field $n_a$ ( $=-D_av$ ), both
introduced in the previous subsection defined on $\ih$ (except the
zero slice of $\ell$). It follows from the $0$th Law \eqref{eq:0th}
that the rotation $1$-form potential is Lie dragged by $\ell$,
\begin{equation}\label{eq:omega_pr}
  \lie_\ell \w{\ell}_a\ =\ 0 \ .
\end{equation}

\subsubsection{Harmonic invariant}
  \label{harmonic}

It turns out, that the rotation 1-form potential $\w{\ell}_a$
defines on the base space $\bas$ a unique harmonic $1$-form depending
only on the geometry $(q_{ab},D_a)$ of $\ih$. Indeed, given the
function $v$, there is a differential 1-form field $\hw{\ell}_A$
defined on $\bas$ and called the projective rotation 1-form potential,
such that
\begin{equation}\label{eq:omega}
  \w{\ell}_a\ =\ \spi^*\hw{\ell}_a\
    +\ \sgr{\ell}D_av \ .
\end{equation}
The 1-form $\hw{\ell}_A$ is not uniquely defined, though. It depends
on the choice of the function $v$ compatible with $\ell^a$, and on the
choice of $\ell^a$ itself. Given $\ell^a$, the freedom is in the
transformations
\begin{subequations}\begin{align}
  v\ &=\ v' + B,\quad \lie_\ell B \ =\ 0\label{eq:v'} \ , \\
  \hw{\ell}'_A \ &=\  \hw{\ell}_A+\sgr{\ell}\hD_A B
  \ .\label{eq:tomega'}
\end{align}\end{subequations}
The transformations $\ell'^a=f\ell^a$ which preserve the condition
(\ref{eq:kappaconst}) are necessarily of the form
\begin{equation}\label{eq:f}
  f\ =\ \begin{cases}
          B e^{-\sgr{\bsl}v} + \fracs{\sgr{\bsl'}}{\sgr{\bsl}}
            & \sgr{\bsl}\neq 0 \\
          \sgr{\bsl'} v - B  &  \sgr{\bsl}= 0 \\
        \end{cases}
\end{equation}
and it can be shown using \eqref{eq:omega'}, that the only possible
form of the corresponding $\hw{\ell'}_A$ is again that of
\eqref{eq:tomega'} with possibly different function $B$ and value
of surface gravity. Therefore, if we apply to $\hw{\ell}_A$ the
(unique) Hodge decomposition onto the exact, the co-exact, and the
harmonic part, respectively,
\begin{equation}\label{eq:Hodge}
  \hw{\ell}{}_A\ =\ \hw{\ell}{}_A^{\rm ex}
  + \hw{\ell}{}_A^{\rm co} + \hw{\ell}{}_A^{\rm ha}\ ,
\end{equation}
then the parts $\hw{\ell}{}_A^{\rm co}$ and
$\hw{\ell}{}_A^{\rm ha}$ are invariant, that is
determined by the geometry $(q_{ab},D_a)$ of $\ih$ only. The
co-exact part is determined by the already defined invariant
2-form (\ref{eq:Omega}), via
\begin{equation}
  \hat{\Omega}_{AB}\ =\ \hD_A \hw{\ell}{}_B^{\rm co}
  - \hD_B \hw{\ell}{}_A^{\rm co}\ .
\end{equation}
The harmonic part of  $\hw{\ell}{}_A$ is the new invariant (see
\cite{lp-g} for details) possibly nontrivial for NEHs of base space
topology different than $S^n$. As the space of harmonic $1$-forms is
finite-dimensional, the degrees of freedom identified with the
harmonic component of the rotation $1$-form potential are global in
the character.

\subsubsection{Jezierski-Kijowski null vector field}

Given a nowhere vanishing null vector field $\ell_o\in\Gamma(T(\ih))$
such that $\sgr{\ell_o}=0$ the rotation $1$-form $\w{\ell_o}$
corresponding to it is a pull-back of the projective rotation $1$-form
$\hw{\ell_o}$. Suppose $\ell'_o=f\ell_o$ is another null vector field
such that its surface gravity \eqref{eq:sgr} vanishes. Then its
rotation $1$-form $\w{\ell'_o}$ is related to $\w{\ell_o}$ via
\eqref{eq:omega'} the
following way:
\begin{equation}\label{eq:JK-w-tr}
  \spi^*\hw{\ell'_o}{}_a\ =\ \spi^*\hw{\ell_o}{}_a + D_a\ln f \ .
\end{equation}
The same transformation rule implies that $\lie_{\ell_o}f=0$, hence
there exists function $\hat{f}:\bas\to\re$ such that $D_a\ln f =
\spi^*(\hD\ln\hat{f})_a$. The equation \eqref{eq:JK-w-tr} can be then
written down as an expression involving objects defined on $\bas$ only
\begin{equation}
  \hw{\ell'_o}_A\ =\ \hw{\ell_o}_A + \hD_A\ln\hat{f} \ .
\end{equation}
In particular $\hat{f}$ can be chosen such that
$\hD_A\ln \hat{f}=-\hw{\ell_o}{}_A^{\rm ex}$ implying
\begin{equation}\label{eq:JK-cond}
  \hw{\ell'_o}{}_A^{\rm ex} = 0 \ .
\end{equation}
Due to the uniqueness of Hodge decomposition the function $\hat{f}$
chosen that way is unique at $\ih$ up to multiplication by a constant,
so is the vector field $\ell_o$ satisfying \eqref{eq:JK-cond}. We will
denote that  null field by $\ellJK$ and refer to it as the {\it
  Jezierski-Kijowski} (J-K) null vector field \cite{JK}.

\subsubsection{Degrees of freedom}
  \label{ssec:degrees}

Let  $\ell^a$, $v$ and $n_a$ be still the same, respectively,
vector field, a compatible coordinate and a covector field
specified at the begin of this subsection. The covariant derivative
$D_a$ is characterized by the elements $\w{\ell},S_{ab}$ (defined in
section \ref{sec:nes-geom}), subject to the constraints
(\ref{eq:[l,D]1}, \ref{eq:[l,D]2}). Suppose the Einstein equations 
with a (possibly zero) cosmological constant are satisfied on $\ih$,
and the field equations of the matter fields possibly present on $\ih$
imply that on each non-expanding surface\footnote{The conditions below
  are satisfied for example by the Maxwell field in $4$-dimensional
  spacetime}
\begin{subequations}\label{eq:mat-cond}\begin{align}
  \lie_{\ell}T_{ab}\ &=\ 0 \ ,  &
  \ell^aT_{ab}\ &= \ 0 \ , \tag{\ref{eq:mat-cond}}
\end{align}\end{subequations}
where $T_{ab}$ is a pull-back to $\ih$ of the matter energy-momentum
tensor.\\
{\it The geometry $(q_{ab},D_a)$ can be completely
characterized by the following data:
\begin{enumerate}[ (i)]
  \item defined on the space of the null geodesics $\bas$:
     \begin{itemize}
       \item the projective metric tensor $\hq_{AB}$
     \eqref{eq:hq}.
       \item the projective rotation 1-form potential
         $\hw{\ell}{}_A$ \eqref{eq:omega}
       \item the projective transversal expansion-shear data
         $\hS^o_{AB}$ (see \eqref{eq:transexpsh} below)
     \end{itemize}
   \item the values of the surface gravity $\sgr{\ell}$ and the
         cosmological constant $\Lambda$,
   \item (in non-vacuum case) the projective matter energy-momentum
     tensor $\hat{T}_{AB}$ defined via $T_{ab}=:(\spi^*\hat{T})_{ab}$,
\end{enumerate}}
where the projective transversal expansion-shear data
$\hS^o_{AB}$ is a tensor defined on $\bas$  by the following form
of a general solution to \eqref{eq:S_ev},
\begin{equation}\label{eq:transexpsh}
  \tS_{AB}\ =\ \begin{cases}
    \begin{split}
      &v\, \tq^{a}{}_{A}\tq^{b}{}_{B}
        \left((\spi^*\hD\hw{\ell})_{(ab)}
          + (\spi^*\hw{\ell})_a(\spi^*\hw{\ell})_a
      + \fracs{1}{2}(\spi^*\hat{T})_{ab}\right)+ \\
      &\ +\, v\ \left(
      - \fracs{1}{2}\,\Ricn{n-2}_{AB}
          - \fracs{1}{2}\Lambda \tq_{AB}
          \right) \
      +\ \tq^{a}{}_{A}\tq^{b}{}_{B}(\spi^*\hS^o)_{ab}
    \end{split} &  \text{for } \ \ \sgr{\ell}=0 \ , \\
    \begin{split}
      &\frac{1}{\sgr{\ell}}\tq^{a}{}_{A}\tq^{b}{}_{B}
        \left((\spi^*\hD_\hw{\ell})_{(ab)}
          + (\spi^*\hw{\ell})_a(\spi^*\hw{\ell})_b
      + \fracs{1}{2}(\spi^*\hat{T})_{ab} \right)+  \\
      &\ +\, \frac{1}{\sgr{\ell}}\left(
        - \fracs{1}{2}\,\Ricn{n-2}{}_{AB}
        - \fracs{1}{2}\Lambda \tq_{AB}\right)\
      +\ e^{-\sgr{\ell} v}\tq^{a}{}_{A}\tq^{b}{}_{B}(\spi^*\hS^o)_{ab}
    \end{split}  & \text{otherwise.}\\
  \end{cases}
\end{equation}
A part of data depends on the choice of the vector field $\ell^a$
and the compatible coordinate $v$. Given $\ell^a$ such that
$\sgr{\ell}\not=0$, the compatible coordinate $v$ can be fixed up
to a constant by requiring that the exact part in the Hodge
decomposition of the projective rotation 1-form potential
$\hw{\ell}{}_A$ vanishes (see subsection \ref{harmonic}). The vector
$\ell^a$ itself, generically, can be fixed up to a constant factor
by requiring that the projective transversal expansion-shear data
$\hS^o{}_{AB}$ be traceless.

Finally, the remaining rescaling freedom by a constant can be removed
by fixing the value of the surface gravity $\sgr{\ell}$ arbitrarily
(the area of $\ih$ can be used as a quantity providing the appropriate
units).

\subsection{Abstract NEH geometry, maximal analytic extension}
  \label{sec:abstract}

\subsubsection{Abstract NES/NEH geometry}

Non-expanding null-surface/horizon geometry can be defined more
abstractly. Consider an $(n-1)$-dimensional manifold $\ih$. Let
$q_{ab}$ be a symmetric tensor of the signature $(0,+...+)$. Let $D_a$
be a covariant, torsion free derivative such that
\begin{equation}\label{eq:Dq-abs}
  D_aq_{bc}\ =\ 0 \ .
\end{equation}
A vector $\ell^a$ tangent to $\ih$ is called null whenever
\begin{equation}\label{eq:lq-abs}
  \ell^aq_{ab}\ =\ 0 \ .
\end{equation}
Even-though we are not assuming any symmetry, every null vector
field $\ell^a$ is a symmetry of $q_{ab}$,
  \begin{equation}
    \lie_\ell q_{ab}\ =\ 0 \ .
  \end{equation}

Given a null vector field $\ell^a$, we can repeat the definitions
of section \ref{sec:nes-def} and associate to it the surface gravity
$\sgr{\ell}$, and the rotation 1-form potential $\w{\ell}$. Now, an
Einstein constraint corresponding to a matter energy-momentum tensor
$T_{ab}$ satisfying \eqref{eq:mat-cond} can be defined as an
equation on the geometry  $(q_{ab},D_a)$ per analogy with the
non-expanding null surface case. To spell it out we need one more
definition. Introduce on $\ih$ a symmetric tensor $\Ricn{n-2}_{ab}$,
such that for every $(n-2)$-subsurface contained in $\ih$ the pullback
of $\Ricn{n-2}_{ab}$ to the subsurface coincides with the Ricci tensor
of the induced metric, provided the induced metric  is
non-degenerate. The constraint is defined as
\begin{equation}\label{eq:einsteinvac}
  [\lie_\ell,D_a]_c^b\ =\ \ell^b\left[ \left(D_{(a}\w{\ell}{}_{c)}\
    +\ \w{\ell}{}_a\w{\ell}{}_c\ -\ \fracs{1}{2}\Lambda q_{ac}\right)\
    -\ \fracs{1}{2}\Ricn{n-2}_{ac}\ +\ \fracs{1}{2}T_{ac}
  \right] \ ,
\end{equation}
(where $T_{ab}$ is a symmetric tensor satisfying \eqref{eq:mat-cond})
and it involves an arbitrary  cosmological constant $\Lambda$.

Suppose now, that
\begin{equation}
  \ih\ =\ \bas\times \mathbb{R}\ ,
\end{equation}
and the tensor $q_{ab}$ is the product tensor defined naturally by a
metric tensor $\hq_{AB}$ defined in $\bas$ and the identically zero
tensor defined in $\mathbb{R}$. The analysis of subsections
\ref{sec:nes-geom}, \ref{sec:neh} can be repeated for solutions of
the Einstein constraint \eqref{eq:einsteinvac}. Again the base space
$\bas$ is equipped with the data specified in section
\ref{ssec:degrees}, that is the projective: metric tensor 
$\hq_{AB}$, rotation 1-form potential $\hw{\ell}_{A}$, transversal
expansion-shear data $\hS^o{}_{AB}$, matter energy-momentum tensor
$\hat{T}_{AB}$. Completed by the values of the surface gravity
$\sgr{\ell}$ and the cosmological constant $\Lambda$ the data is free,
in the sense that every data set defines a single solution
$(q_{ab},D_a)$.

\subsubsection{Maximal analytic extension}
  \label{ssec:ext}

Suppose $\ih$ is an abstract non-expanding horizon described
above. Suppose also that $u$ is an affine parameter defined globally
on $\ih$ and parametrizing its null curves
\begin{equation}\label{eq:u}
  \ell_o^{a}u_{,a}\ =\ 1 \ , \quad \sgr{\ell_o}\ =\ 0 \ .
\end{equation}
Given on the horizon base space any local coordinate system
$(\hat{x}^A) = (\hat{x}^1,\ldots,\hat{x}^{n-2})$ one can define the
coordinate system $(x^a)=(x^A,u)$ at $\ih$ where
\begin{equation}\label{eq:coord}
  x^A\ :=\ \spi^*\hat{x}^A \ ,
\end{equation}
and $\spi$ is a projection defined via \eqref{eq:NEH-pi-def}.
In the corresponding frame $e^a_b = x^a_{,b}$, the coordinate
component of metric tensor $q_{ab}$ are constant along the null curves
whereas the components of $D_a$ are determined by the projective data
via equations\footnote{We still assume that matter fields satisfy the
  conditions \eqref{eq:mat-cond}.} (\ref{eq:omega},
\ref{eq:transexpsh})\footnote{Case $\sgr{\ell}=0$.} (with $v$ in
\eqref{eq:transexpsh} replaced by $u$). Hence all the geometry
components depend analytically on the parameter $u$. Considered
abstract $\ih$ can be then extended in the parameter $u$ such that its
geometry components are determined by the equations (\ref{eq:hq},
\ref{eq:omega}, \ref{eq:transexpsh}) and all the null geodesics are
complete. Such extension is regular due to finiteness of the solutions
to the system (\ref{eq:hq}, \ref{eq:omega},\ref{eq:transexpsh}) for
finite $u$. Also, as every two affine parameters $u,u'$ correspond to
each other the following way
\begin{equation}
  u'\ =\ au + b \ ,
\end{equation}
(where $a$, $b$ are constant along the null geodesics) the analyticity
and the extension of $(\ih,q_{ab},D_a)$ do not depend on a choice
of an affine parameter on $\ih$.
 We will denote this extension by $\eih$ and refer to
it as the maximal extension of a non-expanding horizon/null
surface (MAENEH/MAENES). The coordinate system $(x^A,u)$ defined via
(\eqref{eq:u}, \ref{eq:coord}) extends straightforward onto $\eih$.

\section{Symmetries: Definition and basic properties}
  \label{sec:symm-basic}

\subsection{The definitions}
  \label{sec:symm-def}

\begin{defn}\label{def:symm}
  Given a non-expanding horizon $\ih$ of the induced metric $q_{ab}$
  and covariant derivative $D_a$, a vector field $X\in\Gamma(T(\ih))$
  will be called an infinitesimal symmetry if
  \begin{equation}\label{eq:symm_def}
    \lie_{X}q_{ab}\ =\ 0 \quad\text{and}\quad
    [\lie_X,D_a]\ =\ 0        \ .
  \end{equation}
\end{defn}

A non-expanding horizon $\ih$ admitting an infinitesimal symmetry $X$
will be referred to as {\it symmetric}.
An example of a symmetric NEH is an Isolated Horizon \cite{abl-g,lp-g},
that is a NEH which admits  a null infinitesimal symmetry additionally
assumed to be nowhere vanishing.

A (locally defined) diffeomorphism $U:\ih\rightarrow \ih$
is called a (local) {\it symmetry} of a NEH $\ih$ if it preserves
the horizon geometry $(q_{ab},D_a)$.

Given an infinitesimal symmetry $X$ of a NEH, consider the
corresponding local diffeomorphism flow $U_t$, that is a family of
local diffeomorphisms labeled by a real parameter $t$, such that
\begin{subequations}\label{eq:symm-diff}\begin{align}
  U_t\circ U_{t'}\ &=\ U_{t+t'} \ , &
  U_0 \ &= \ \Id \ , &
  X^aD_a f \ &=\ \frac{d}{dt}|_{t=0}U_t^*f \ ,
  \tag{\ref{eq:symm-diff}}
\end{align}\end{subequations}
for every function $f$.  The flow preserves all the structures
defined by the horizon geometry. For example a function
$u:\ih\rightarrow \re$ such that restricted to every null geodesic
curve in $\ih$ becomes an affine parametrization -- we refer to $u$
briefly as {\it an affine parameter on $\ih$} -- is mapped by the
pull back into a locally defined affine parameter on $\ih$. The
properties implied by the preservation of the horizon geometry are
enlisted below in the following:
\begin{cor}\label{cor:NEH-sym-f}
  Suppose $X$ is an infinitesimal symmetry of a horizon, and $U_t$ is
  the corresponding local diffeomorphism  flow. Then the following is
  true:
  \begin{enumerate}[ (i)]
    \litem For every null vector field $\ell\in\Gamma(T(\ih))$
      \begin{equation}
        \lie_X\ell^a\ =\ a\ell^a \ ,
      \end{equation}
       where $a$ is a function depending on $\ell$,\label{it:csf-dir}
    \litem provided the surface gravity $\sgr{\ell}$ of the
    vector field $\ell$ is constant on $\ih$,
      \begin{equation}
         \sgr{U_{t*}\ell}\ =\ \sgr{\ell} \ ,
      \end{equation} \label{it:csf-act}
    \litem the Jezierski-Kijowski vector $\ellJK$ is preserved up to
      a multiplicative constant
      \begin{equation}\label{eq:symm-JK-comm}
        \lie_X\ellJK\ =\ -\sgr{X}\ellJK\ , \quad \sgr{X}\ =\ \const\ ,
      \end{equation} \label{it:csf-JK}
    \litem every affine parameter $u$ on
    $\ih$ is mapped by the pull back in the following way
      \begin{equation}\label{eq:u-tr}
        U_t^*u\ =\ au + b \ ,
      \end{equation}
      where $a,b$ are functions constant along null geodesics at
      $\ih$. In particular, the parameter $\bar{v}$ compatible with
      J-K vector $\ellJK$ transforms upon the action $U_t$ as follows
      \begin{equation}\label{eq:JK-u-tr}
        U_t^*\bar{v}\ =\ e^{\sgr{X}t}\bar{v} + b \ ,
      \end{equation}
      where $b$ is constant along null geodesics and the constant
      $\sgr{X}$ is defined in (\ref{eq:symm-JK-comm}).\label{it:csf-u}
  \end{enumerate}
\end{cor}
\begin{proof}
Indeed, the point \eqref{it:csf-dir} follows from the first
equation in definition \ref{eq:symm_def} and from the fact that
$\ell$ is tangent to the unique degenerate direction of $q_{ab}$.

The point $(ii)$ follows from the following equation
\begin{equation}
(U_{t*}\ell^a)D_aU_{t*}\ell\ =\ U_{t*}(\sgr{\ell}\ell)\ =\
\sgr{\ell}U_{t*}\ell\ =:\ \sgr{U_{t*}\ell}U_{t*}\ell \ .
\end{equation}

The uniqueness (up to multiplicative constant) of J-K
vector implies \eqref{it:csf-JK}.

The transformation law \eqref{eq:JK-u-tr} follows from the
integration of \eqref{eq:symm-JK-comm}.
\end{proof}

The constant $\sgr{X}$ assigned to an infinitesimal symmetry $X$
in (\ref{eq:symm-JK-comm}) will play an important role in the
analysis of the symmetric NEHs. Some of our conclusions will
be sensitive on the vanishing of  $\sgr{X}$:
\begin{defn}\label{def:extrsym}
  An infinitesimal symmetry $X$ of a NEH is called extremal if
  \begin{equation}
    \sgr{X}=0.
  \end{equation}
  Otherwise $X$ is called a non-extremal infinitesimal symmetry.
\end{defn}
Note that the symmetry group generated by an extremal infinitesimal
symmetry preserves the Jezierski-Kijowski vector.

As it was explained in the previous section, we will consider in
this paper those NEHs whose geometry is analytic in (any and
each) affine parameter $u$ defined on $\ih$. That assumption is
enforced by the Einstein equations for vacuum or a large class of
matter fields. The analyticity leaded to the definition of the
maximal analytic extension of a given NEH $\ih$ and its geometry.
Since the extension exists and is unique, it can be always taken.
Therefore, given an infinitesimal symmetry of a NEH, it is natural to
ask, whether it is also analytic in an affine parameter. The
answer is in the affirmative:

\begin{prop} \label{prop:symm-ext}
  Suppose $\ih$ is a NEH and $X$ is an infinitesimal symmetry. Then:
  \begin{enumerate}[ (i)]
    \litem in every local coordinate system of the
      form (\ref{eq:u}, \ref{eq:coord}) $X$ is analytic in the affine
      parameter $u$,
    \litem there is uniquely defined analytic extension
      $\bar{X}$ of $X$  to the maximal analytic extension $\eih$
    \litem $\bar{X}$ is an infinitesimal symmetry of the MAENEH $\eih$,
    \litem  $\bar{X}$ generates a group of globally defined
      symmetries of $\bar{\ih}$.
  \end{enumerate}
\end{prop}

\begin{proof} Consider a coordinate system $(x^1,...,x^{N-2},u)=(x^A,u)$
defined in (\ref{eq:u}, \ref{eq:coord}). We have
\begin{equation}
  X\ =\ X^A\partial_A + X^u\partial_u
\end{equation}
It follows from corollary \ref{cor:NEH-sym-f}\eqref{it:csf-dir} that
\begin{equation}
  \partial_uX^A\ =\ 0 \ ,
\end{equation}
hence they are analytic in $u$ in the trivial way. The function
$X^u$, on the other hand,  according to \eqref{eq:u-tr} is at most
linear in $u$. Therefore indeed, $X$ is analytic in $u$ and
extendable to the maximal analytic extension $\eih$. Via the
analyticity it continues to be the extended infinitesimal symmetry
of the horizon. Therefore, we may assume that $\ih$ is maximal in the
sense it equals its maximal analytic extension. Given an affine
parameter $u$ on $\ih$, there is defined a diffeomorphism
\begin{equation}
  \ih\ \rightarrow\ \bas\times \re \ , \quad
  p\mapsto (\spi(p),u(p)) \ .
\end{equation}
The local diffeomorphism flow $U_t$ of the vector field $X$
considered in corollary \ref{cor:NEH-sym-f} can be defined for
every compact subset of the horizon $\ih$, provided $t$ is
adjusted appropriately to the subset.  Note first, that due to
corollary \ref{cor:NEH-sym-f} \ref{it:csf-dir}, there is a vector
field $\hat{X}\in\Gamma(T(\bas))$ uniquely defined by the projections
of $X$,
\begin{equation}
  \hat{X}\ =\ \spi_*X \ .
\end{equation}
Because $\bas$ is compact, the  flow of the vector field $\hat{X}$
is a 1-dimensional group of globally defined diffeomorphisms
$\hat{U}_{t'}:\hat{\ih}\rightarrow\ih$. Now, the action of the
flow $U_t$ in $\ih=\bas\times\re $ has the form
\begin{equation}
  U_t(\hat{p},u)\ =\ (\hat{U}_t(\hat{p}),V_{t,\hat{p}}(u)) \ ,
\end{equation}
where the $\hat{U}_t$ is the diffeomorphism flow of $\hat{X}$
independent of $u$. As far as the function $V_{t,\hat{p}}$ is
concerned, it follows from \eqref{eq:u-tr}, it is at most linear
(that is affine). Therefore, given value $t$ it is defined on the
entire $\ih$, and, in the consequence, it is well defined for
every value of $t$.
\end{proof}

An important technical consequence of the existence of the globally
defined flow of the infinitesimal symmetry $\bar{X}$ of the
maximal analytic extension $\eih$, is that the function $b$ defined
in \eqref{eq:JK-u-tr} (depending on the label $t$) is globally defined
on $\ih$, and in fact
\begin{equation}
  b\ =\ \spi^*\hat{b}
\end{equation}
for some globally defined function $\hat{b}:\bas\rightarrow \re$.

Proposition \ref{prop:symm-ext} together with the results of
subsection \ref{sec:abstract} imply that 
in the analysis of symmetries one can then directly use the extended
objects defined on MAENEH $\eih$. Therefore:
\begin{rem}
  From now on we will identify the NEH $\ih$ (and objects defined on
  it) with its analytic extension $\eih$ (and extensions of objects
  defined on $\ih$ respectively), thus dropping the 'bars' in the
  notation.
\end{rem}

\subsection{Symmetry induced on the base space}
  \label{sec:symm-ind}

For the existence of a not everywhere null infinitesimal symmetry
of a NEH $\ih$, necessary conditions have to be satisfied by the
Riemannian geometry of the horizon base space $\bas$. It follows from
corollary \ref{cor:NEH-sym-f} that the projection of $X$ onto horizon
base space $\bas$ defines a unique vector field $\hat{X}\in T(\bas)$.

The equation constituted by the pull-back of (\ref{eq:symm_def}a) onto
$\bas$ yields that the projected field satisfies the condition
\begin{equation}
  \lie_{\hat{X}}\hq_{AB} = 0 \ .
\end{equation}
Hence non-null symmetry of the horizon generates a symmetry of
$\hq_{AB}$. The field $\hat{X}$ will be referred to as the {\it
  infinitesimal symmetry induced} by $X$. It generates a flow $\re\ni
t\mapsto \hat{U}_t$ of globally defined isometries of $\bas$.

Those properties allows us to classify the non-null horizon symmetries
with respect to the classification of the Killing fields of the
compact Riemann geometry of $\bas$. In particular the
classification of symmetric geometries defined on $S^2$ \cite{abl-m}
is applied in section \ref{sec:4D-class} where complete classification
of the symmetric $3$-dimensional NEHs (in $4$-dimensional spacetime)
is presented.

\section{Null symmetries}
  \label{sec:symm-gen-null}

\begin{defn}\label{def:null-def}
  If an infinitesimal symmetry $X$ of a NEH $\ih$ is a null vector
  field, then:
  \begin{enumerate}[ (i)]
    \litem $X$ is called a null infinitesimal symmetry,
    \litem the symmetry group generated by $X$ on $\ih$
      is called a null symmetry group.
  \end{enumerate}
\end{defn}
A NEH admitting an infinitesimal null symmetry (denoted here by $\bsl$)
will be referred to as a {\it null symmetric} horizon. An important
class among such horizons is constituted by {\it isolated horizons}
(IH): the ones whose null infinitesimal symmetries nowhere vanish. The
detailed analysis of their geometry can be found in \cite{abl-g,lp-g}.
The general null symmetry case, however, requires more care, because,
unlike in the IH case, now we {\it do not assume $\bsl$ does not
  vanish}.

\subsection{Non-extremal null symmetry in arbitrary dimension}
  \label{sec:nullsym-nex}

\begin{theorem}\label{thm:ell-nex}
  Suppose $\bsl$ is a null non-extremal infinitesimal symmetry of a
  NEH $\ih$.
  Then the zero set of $\bsl$ is a global section of the projection
  $\spi:\ih\rightarrow\bas$, provided the Stronger Energy Condition
  \ref{c:energy} holds for matter fields present on $\ih$.
\end{theorem}

\begin{proof}
Fix on $\ih$ a null vector field  $\ell_o$ such that
$\sgr{\ell_o}=0$ and a coordinate $v$ compatible with it.

The non-extremal null infinitesimal symmetry $\bsl$ on $\ih$
can be expressed by $\ell_o$ as
\begin{equation}
  \bsl\ =\ f\ell_o \ ,
\end{equation}
where the form of the proportionality coefficient
$f$ is determined by \eqref{eq:f}
\begin{equation}\label{eq:nex-sym-form}
  f\ =\ \sgr{\bsl}v + B \ ,
\end{equation}
and $B$ is a real function defined on the entire $\ih$.
Now, the zero set of $\bsl$ is given by the equation
\begin{equation}
  v\ =\ -\frac{B}{\sgr{\bsl}}\ =\ 0 \ .
\end{equation}
\end{proof}
The zero set of a non-extremal null infinitesimal symmetry
is called its cross-over surface.

\subsection{Extremal null symmetry in arbitrary dimension}
  \label{sec:nullsym-ex}

\begin{theorem}\label{thm:ell-ex}
 Suppose a NEH $\ih$ admits a null and extremal infinitesimal symmetry
 $\bsl$. Suppose also that the pull-back $T_{ab}$ onto $\ih$ of the
 energy-momentum tensor of the matter fields possibly present on
 $\ih$ satisfies the condition \ref{eq:mat-cond}.
 Then the following holds:
  \begin{itemize}
    \litem The infinitesimal symmetry $\bsl$ doesn't vanish at a(n
      open and) dense  subset of $\eih$.
    \litem If $\bsl$ vanishes at some point $p\in\ih$ it also vanishes
      at the entire null geodesics intersecting $p$. The set of null
      geodesics on which $\bsl=0$ forms in $\bas$ a surface defined by
      the equation
      \begin{equation}
        \hat{B}\ =\ 0 \ ,
      \end{equation}
      where $\hat{B}$ is a real valued function defined on $\bas$ such
      that for every $\hat{p}\in\bas$
      \begin{equation}
        \hat{B}(\hat{p})\ =\ 0\quad\Rightarrow\quad
        \hd\hat{B}(p)\ \neq\ 0 \ .
      \end{equation}
  \end{itemize}
\end{theorem}

\begin{proof}
Let $\ell_o$ be  a globally defined on $\ih$  null vector field
tangent to $\ih$ and such that $\sgr{\ell_o}=0$. Due to \eqref{eq:f}
there is a function $\hat{B}$ defined on $\bas$ such that
\begin{equation}\label{eq:ex-symm-JK}
  \bsl\ =\ B\ell_o \ , \quad B\ =\ \spi^*\hat{B} \ .
\end{equation}
In consequence, if $\bsl$ vanishes at some point $p\in\ih$ it also
vanishes at the entire null geodesics $\hat{p}$.

Denote the set formed by geodesics on which $\hat{B}\neq 0$ by
$\hat{U}\subset\bas$.
As $\bsl$ is a null infinitesimal symmetry, the Lie derivative
$N'{}_{ab}$ of the transversal expansion-shear tensor corresponding to
the coordinate compatible with $\bsl$ vanishes (see (\ref{eq:[l,D]1},
\ref{eq:com_form}, \ref{eq:symm_def}b)) everywhere where $\bsl\neq
0$. That leads to the following constraint on $B$
\begin{equation}\label{eq:N-trans}
  BN_{bc} + \w{\bsl}_cD_bB + \w{\bsl}_bD_cB + D_bD_cB\
  =\ BN'{}_{bc}\ =\ 0 \ ,
\end{equation}
where $N_{ab}:=\lie_{\ell_o}S_{ab}$ with $S_{ab}$ (defined via
\eqref{eq:S_def}) being associated with the coordinate $v$
compatible with $\ell_o$.

Projecting this equation onto surfaces $v=\const$ and expressing the
projected objects by projective data via \eqref{eq:transexpsh} we find
that $\hat{B}$ satisfies the following PDE
\begin{equation}\label{eq:nullsym-full}
  \left[ \hD_A\hD_B + 2\hw{\ell^o}_{(A}\hD_{B)}
         + (\hD_{(A}\hw{\ell^o}_{B)})
         + \hw{\ell^o}_A\hw{\ell^o}_B - \fracs{1}{2}\Ricn{n-2}_{AB}
         - \fracs{1}{2}\Lambda\hq_{AB} + \fracs{1}{2}\hat{T}_{AB}
  \right]\hat{B} \ = \ 0 \ ,
\end{equation}
(where $T_{ab} = \spi^*\hat{T}_{ab}$) for the function $\hat{B}$ on
$\bas$.

Note that the above equation holds on the entire $\bas$ as it is
satisfied on the closure of the subset such that $\hat{B}\not=0$ on
the one hand, and on the other hand it holds trivially on the
remaining open subset since $\hat{B}=0=\hD_A\hat{B}=\hD_A\hD_B\hat{B}$
therein.

For every solution $\hat{B}$ to this equation the following is true
\begin{lem}\label{lem:nullsym-diff}
  Suppose $\hat{B}$ is  a solution to the equation
  \eqref{eq:nullsym-full}. If $\hat{B}(p) = d\hat{B}(p) = 0$ at some
  point $p\in\bas$ then $\hat{B}$ vanishes on the entire $\bas$.
\end{lem}

\begin{proof}[Proof of the Lemma \ref{lem:nullsym-diff}]
Consider on $\bas$ a geodesics $\hat{\gamma}$ parametrized by an
affine parameter $x$. Taking the (double) contraction of the equation
\eqref{eq:nullsym-full} with the vector $\dot{\gamma}^A$ tangent to
$\hat{\gamma}$ we obtain the following constraint for the value of
function $\hat{B}$ at $\hat{\gamma}$
\begin{equation}
  \left[ \fracs{\rd^2}{\rd x^2}\
         +\ 2\hw{\ell_o}{}_{\hat{X}}\fracs{\rd}{\rd x}\
         +\ (\fracs{\rd}{\rd x}\hw{\ell_o}{}_{\dot{\gamma}})\
         +\ \hw{\ell_o}{}^2_{\dot{\gamma}}
         - \fracs{1}{2}\hRicn{n-2}_{\dot{\gamma}\dot{\gamma}}
         - \fracs{1}{2}\Lambda|\dot{\gamma}^A|^2_{\hq}
         + \fracs{1}{2}\hat{T}_{\dot{\gamma}\dot{\gamma}}
  \right] \hat{B}\ =\ 0 \ ,
\end{equation}
where $\hw{\ell_o}_{\dot{\gamma}}:=\hw{\ell_o}_A\dot{\gamma}^A$,
$\hRicn{n-2}_{\dot{\gamma}\dot{\gamma}} :=
\hRicn{n-2}_{AB}\dot{\gamma}^A\dot{\gamma}^B$ and
$\hat{T}_{\dot{\gamma}\dot{\gamma}} :=
\hat{T}_{AB}\dot{\gamma}^A\dot{\gamma}^B$.

This equation constitutes a linear homogeneous ODE for $\hat{B}$. We
assume that the geometry of considered NEH is regular so is projective
geometry induced on $\bas$. Thus the equation satisfies Lipschitz
rule. In consequence the value of $\hat{B}$ at the entire
$\hat{\gamma}$ is determined by initial values $\hat{B}$ and
$\fracs{\rd}{\rd x}\hat{B}$ at some starting point
$p\in\hat{\gamma}$.

Suppose now, that there exists on $\bas$ the point $p_o$ such that
$\hat{B}|_{p_o}=\hd\hat{B}|_{p_o}=0$. Then on every geodesics
$\hat{\gamma}$ intersecting $p_o$ the function $\hat{B}$ vanishes as
$\hat{B}=0$ is an unique solution to the initial value problem
$\hat{B}|_{p_o}=\fracs{\rd}{\rd x}\hat{B}|_{p_o}=0$. On the other hand
every two points on the connected manifold can be connected via
geodesics, hence $\hat{B}$ vanishes on the entire $\bas$.
\end{proof}

From the Lemma \ref{lem:nullsym-diff} follows immediately that the
gradient of $\hat{B}$ cannot vanish at the point on which
$\hat{B}=0$.
\end{proof}

\subsection{Higher-dimensional null symmetry group}
  \label{sec:nullsym-general}

Given a NEH(MAENEH) $\ih$ admitting an infinitesimal null
symmetry $\bsl$ there may exist another, linearly independent
null infinitesimal symmetry $\bsl'$ . The
consequences of the existence of two null symmetries in the case both
of them nowhere vanish have been considered (for arbitrary dimension
and base space topology) in \cite{lp-g}. The theorems
\ref{thm:ell-nex} and \ref{thm:ell-ex} however allow the
straightforward generalization of those results: as for given
infinitesimal null symmetry $\bsl$ of a NEH $\ih$ the subset of $\ih$
on which $\bsl\neq 0$ is dense in $\ih$ the analysis done in
\cite{lp-g} can be repeated directly for arbitrary null infinitesimal
symmetry. Thus the following is true:
\begin{theorem}\label{thm:nullsym-ndim}
  Suppose $\ih$ is a NEH admitting two distinct infinitesimal null
  symmetries. Suppose also the Stronger Energy Condition
  \eqref{c:energy} holds on $\ih$. Then $\ih$ admits also an extremal
  (see Def. \ref{def:extrsym}) infinitesimal null symmetry.
\end{theorem}

\section{Cyclic and axial symmetries}
  \label{sec:symm-gen-axi}

\subsection{Definition, preferred slices}

\begin{defn}\label{def:axi-def}
  Given a NEH $\ih$ a vector field $\Phi^a\in T\ih$ is cyclic
  infinitesimal symmetry whenever the following holds:
  \begin{itemize}
    \litem $\Phi^a$ is an infinitesimal symmetry of $\ih$ (satisfies
      the equations \eqref{def:symm}),
    \litem the symmetry group of $\ih$ it generates  is diffeomorphic
      to $SO(2)$,
    \litem $\Phi^a$ is spacelike at the points it doesn't
      vanish.
  \end{itemize}
\end{defn}

We will be assuming an infinitesimal cyclic symmetry is normalized,
such that the flow $\re\ni\varphi\mapsto U_\varphi$ it generates
has the period $2\pi$.

A NEH admitting an infinitesimal cyclic symmetry will be referred
to as the {\it circular} horizon.

If the group of the symmetries of $\ih$  generated by a cyclic
infinitesimal symmetry has a fixed point, then we call the
infinitesimal symmetry {\it axial}. A NEH admitting such a symmetry
will be then called an {\it axial} horizon.

In this subsection we study circular NEHs of arbitrary dimension and
topology. All the statements made here apply in particular to
axial horizons. Of course all the circular horizons such that
$\bas=S^2$ are necessary axial, but when dealing with general
horizons we need to relax the axis existence assumption. An event
horizon (of the base space topology $S^2\times S^1$) admitted by a
spacetime described in \cite{ring} is a good example of circular (and
not axial) NEH as it admits the symmetry induced by axial
Killing field which has no fixed points at the horizon.

For every cyclic (axial) infinitesimal symmetry  $\Phi^a$ the
corresponding projective field $\hat{\Phi}^A=\spi_*\Phi^a$ also
generates an isometric action of SO(2) on $\bas$ (which has a fixed
point). The flow will be denoted by $\re\ni\varphi\mapsto
\hat{U}_\varphi$. The integral lines  of both $\Phi^a$ at $\ih$ and
$\hat{\Phi}^A$ at $\bas$ are closed.

For every circular NEH $\ih$, the  cyclic infinitesimal symmetry
$\Phi^a$ is extremal:
\begin{lem}\label{lem:l0-phi-comm}
  Suppose $\ih$ is a circular NEH and $\Phi^a$ is its cyclic
  infinitesimal symmetry. Then $\Phi^a$ commutes with the
  Jezierski-Kijowski vector field,
  \begin{equation}\label{eq:[PhiJK]}
    \lie_{\Phi}\ellJK = 0 \ .
  \end{equation}
  Moreover, there is on $\ih$ a coordinate $u$ compatible with
  $\ellJK$ such that
  \begin{equation}\label{eq:JKu}
    \ellJK^aD_au\ =\ 0.
  \end{equation}
\end{lem}
\begin{proof}
Indeed elements of the symmetry group generated by the cyclic
infinitesimal symmetry $\Phi^a$ can be labeled by a parameter,
$[0,2\pi]\ni \varphi \mapsto\ U_\varphi$ such that
\begin{subequations}\label{eq:_ax}\begin{align}
  U_{\varphi_1}\circ U_{\varphi_2}\ &=\ U_{\varphi_1+\varphi_2} \ , &
  U_0=U_{2\pi} &= \Id \ .  \tag{\ref{eq:_ax}}
\end{align}\end{subequations}
On the other hand the transformation of $\ellJK$ upon $U_\varphi$ is
determined by \eqref{eq:JK-u-tr}
\begin{equation}\label{eq:axi-act-ell}
  {U_\varphi}_*\ellJK\ =\ e^{\sgr{\Phi}\varphi}\ellJK \ .
\end{equation}
Therefore $\sgr{\Phi}$ vanishes and $\ellJK$ is invariant.

Let $u':\ih\rightarrow \re$ be any coordinate compatible with
$\ellJK$. Consider the average over a cyclic symmetry group
\begin{equation}\label{eq:v-average}
  u(p) := \frac{1}{2\pi}\int_{0}^{2\pi}
    u'(U_{\varphi}(p))\rd\varphi.
\end{equation}
It follows from \eqref{eq:[PhiJK]} that
\begin{equation}
  \ellJK^aD_au\ =\ 1 \ .
\end{equation}
The condition \ref{eq:JKu} follows from the invariance of the measure
$d\varphi$. Also, the function $u$ is as many times differentiable as
the integrand $u'$.
\end{proof}

We can summarize the observations made in this subsection by the
following:
\begin{cor}\label{cor:axi-form}
  Suppose a non-expanding horizon $\ih$ admits a cyclic infinitesimal
  symmetry $\Phi^a$. Suppose also that  $\ell\in\Gamma(L)$ is such
  that
  \begin{subequations}\label{eq:c-l}\begin{align}
    \sgr{\ell}\ &=\ \const\ ,  &
    [\Phi,\ell]\ &=\ 0\ ,  \tag{\ref{eq:c-l}}
  \end{align}\end{subequations}
  Then, there exists a diffeomorphism
  \begin{equation}
    h:\ih\ \rightarrow\ \bas\times\re
  \end{equation}
  such that
  \begin{equation}
    h_*\Phi\ =\ (\hat{\Phi},0) \ ,\quad
    h_*\ell\ =\ (0,\partial_u) \ ,
  \end{equation}
  where $\hat{\Phi}=\spi_*\Phi$.
  In particular, this is true for $\ell=\ellJK$.
\end{cor}

\subsection{Cyclic symmetry and null symmetry}

Consider a NEH $\ih$ which admits two infinitesimal symmetries, a
cyclic one $\Phi$ and a null one $\bsl$.

If $\ih$ admits exactly one (modulo a rescaling) null infinitesimal
symmetry or if $\ih$ admits exactly one (modulo a rescaling)
{\it extremal} null infinitesimal symmetry (and possibly other null
infinitesimal symmetries) then arguments similar to those used in the
proof of Lemma \ref{lem:l0-phi-comm} show that $\Phi$ necessarily
commutes with the vector field $\bsl$. In general, when no uniqueness
is assumed, the following can be shown:
\begin{cor}\label{cor:IH-axi-comm}
  Suppose a non-expanding horizon $\ih$ admits a cyclic infinitesimal
  symmetry $\Phi$ and a non-extremal null infinitesimal symmetry
  $\bsl'$. Then, $\ih$ admits a null non-extremal infinitesimal
  symmetry $\bsl$ commuting with $\Phi$,
  \begin{equation}
    [\Phi,\bsl]\ =\ 0 \ .
  \end{equation}
\end{cor}
\begin{proof}
Consider the group average
\begin{equation}
  \bsl\ =\ \int_0^{2\pi} d\varphi {U_\varphi}_*\bsl' \ .
\end{equation}
The resulting vector field necessarily commutes with $\Phi$.
It is also a null infinitesimal symmetry {\it or} it is identically
zero. However, since the symmetry $\bsl'$ is non-extremal, according
to the Theorem \ref{thm:ell-nex} it vanishes on a single slice (the
cross-over surface) of $\ih$ only.
Both the cross-over surface and cyclic symmetry group are compact,
hence the segment of $\ih$ formed by the orbits of $\Phi$ intersecting
the cross-over surface is also compact. Therefore there exists on
$\ih$ an open set such that $\bsl$ doesn't vanish on it.
As the surface gravity of $\bsl'$ is preserved by every symmetry
${U_\varphi}$ (and in the consequence by the group averaging) $\bsl$
is also non-extremal, hence it vanishes only at a single slice of
$\ih$.
\end{proof}

Our conclusions do not apply to the case of a NEH which admits {\it
  two distinct} symmetry groups each generated by a null extremal
infinitesimal symmetry.

\section{Helical symmetry}
  \label{sec:symm-gen-chir}

As in section \ref{sec:symm-gen-chir} the studies here are
general. We consider here an $n\ge 3$ spacetime case and
maximal analytically extended NEHs (MAENEHs).

\subsection{Definition}

\begin{defn}\label{def:hel-def}
  An infinitesimal symmetry $X^a$ of a NEH $\ih$  is called helical if
  \begin{itemize}
    \litem The symmetry group generated by the projection $\hat{X}^A$
      of $X^a$ onto the base space $\bas$ is diffeomorphic to $SO(2)$,
    \litem there exists an orbit of the symmetry group generated by
      $X^a$ in $\ih$ which is not closed (i.e. diffeomorphic to
      a line).
  \end{itemize}
  A NEH admitting a helical infinitesimal symmetry will be called
  helical.
\end{defn}
We will be assuming that each considered helical infinitesimal
symmetry is normalized such that the isometry flow $\re \ni \varphi
\mapsto \hat{U}_\varphi$ generated by the projection $\hat{X}^A$ has
the period $2\pi$.

As we will see below, the presence of a $1$-dimensional helical
symmetry and the constraints imply more symmetries. In fact, every
horizon admitting the helical symmetry admits also a
cyclic and null symmetry.
The proof of this statement will be divided onto few steps:
\begin{itemize}
  \item First we will construct some uniquely defined null vector
    field tangent to $\ih$ and  commuting with the helical
    infinitesimal symmetry. Since the construction is a generalization
    of Hawking's proof of the BH Rigidity Theorem, we name our vector
    field after Hawking. We will treat separately the two cases:
    'extremal' and 'non-extremal' helical infinitesimal symmetry
    (see definition \ref{def:extrsym}).
  \item Next it will be shown that the Hawking vector is a null
    infinitesimal symmetry. In the consequence, the difference between
    the helical infinitesimal symmetry and the corresponding Hawking
    vector is a cyclic infinitesimal symmetry.
 \end{itemize}

Once proved, the existence of the null infinitesimal symmetry will
ensure via theorems \ref{thm:ell-nex}, \ref{thm:ell-ex} that the set
of open orbits of $X$ is a dense subset of $\ih$.

We will see in the next section, that every symmetric electrovac NEH
of the topology $S^2\times\re$ is either null symmetric, or axial,
or helical. Also, in that case the null symmetry (see theorems
\ref{thm:ell-nex} and \ref{thm:ell-ex-evac}) can vanish only at a
single slice of $\ih$, hence every helical NEH in that case either
is an axial isolated horizon or consists of two IHs separated by
cross-over surface (see theorem \ref{thm:EV-rigid}).

\subsubsection{Helical symmetry general properties}

Consider a helical NEH $\ih$. It is equipped with the
Jezierski-Kijowski null field $\ellJK$ (see the subsection
\ref{sec:neh}). Let $\bar{v}$ be a coordinate compatible with
$\ellJK$. The commutator of $\ellJK$ and $X$ is determined via
\eqref{eq:symm-JK-comm} by the global constant $\sgr{X}$ of the
horizon. It will be shown that presence of a non-extremal (extremal)
helical symmetry imposes the presence of a non-extremal (extremal)
null symmetry.

We are assuming $X$ is  normalized such that  the symmetry flow
$\re\ni\phi\mapsto U_\phi$ of  ${X}$ induces in $\bas$ an isometry
flow $\re\ni\phi\mapsto \hat{U}_\phi$ of the period $2\pi$.
With the group parametrization set as above an action of considered
symmetry on $\ellJK$ rescales $\ellJK$ (due to \eqref{eq:JK-u-tr}) as
follows:
\begin{equation}\label{eq:hel-l0-act}
  U_{\phi*}\ellJK = e^{\sgr{X}\phi}\ellJK \ .
\end{equation}

An action $U_{2\pi}$ preserves every null geodesic curve. Therefore we
can define at $\ih$ the function $s(p)$ such that to each point $p$ it
assigns the 'jump value' corresponding to an action $U_{2\pi}$ of a
helical symmetry
\begin{equation}\label{eq:jump-def}
  s(p) := \bar{v}(U_{2\pi}(p)) - \bar{v}(p) \ .
\end{equation}
The 'jump function' $s$ is defined globally at $\ih$ and
differentiable as many times as the symmetry generator $X$.

\subsection{The Hawking vector field}
\label{sec:hawking}

\begin{defn} \label{def:hawking}
  A Hawking vector field corresponding
  to a helical infinitesimal symmetry $X$ of a NEH $\ih$
  is a null vector field $\ell_{(X)}\in \Gamma(T(\ih))$
  of the following properties:
  \begin{enumerate}[ (a)]
    \item The surface gravity $\sgr{\ell_{(X)}}$  of $\ell_{(X)}$ is
      constant on $\ih$,
      \label{it:Hawk-k}
    \item $\hphantom{.}$\vspace{-0.75cm}
      \begin{equation}\label{eq:Hawk-comm}
	[\ell_{(X)},X]\ =\ 0 \ .
      \end{equation}
    \item Every maximal integral curve of the vector field
      $X-\ell_{(X)}$ is closed (diffeomorphic to $S^1$).
      \label{it:Hawk-orb}
  \end{enumerate}
\end{defn}
Below we will construct a Hawking vector field, given a helical
infinitesimal symmetry.                                       .

First, we will establish a certain necessary condition for a null
vector field to be the Hawking one.

In terms of the Jezierski-Kijowski vector field and coordinate
$\bar{v}$ compatible to it, a Hawking field $\ell_{(X)}$ (if it
exists) is of the following form
\begin{equation}\label{eq:lXlJK}
  \ell_{(X)}\ =\ (\sgr{\ell_{(X)}} \bar{v} + b)\ellJK \ ,
\end{equation}
where $b$ is a function defined globally on $\ih$, constant along each
null curve in $\ih$. It follows from the assumed commuting of
$\ell_{(X)}$ with $X$, corollary \ref{cor:NEH-sym-f} \eqref{it:csf-JK}
and \eqref{eq:u-tr}, that
\begin{equation}\label{eq:sgr-ell-X}
  \sgr{\ell_{(X)}}\ =\ \sgr{X} \ .
\end{equation}
Moreover there exists choice of the coordinate $\bar{v}$ compatible
with $\ellJK$ such that
\begin{equation}
  X^aD_ab\ =\ 0 \ .
\end{equation}

A Hawking vector field  $\ell_{(X)}$ will be found
for each case (non-extremal and extremal) independently. Also the
(more general) method of systematic derivation of it will be presented
in appendix \ref{ssec:Hawk-syst}.
Let us start with the case of extremal $X$ first.

\subsubsection{Extremal case}

Suppose the helical infinitesimal symmetry $X$ is extremal,
\begin{equation}
  \sgr{X}\ =\ 0 \ .
\end{equation}
We will show that the vector field
\begin{equation}
  \ell_{(X)}\ =\ \frac{s}{2\pi}\ellJK \ ,
\end{equation}
is a corresponding Hawking vector, where $s$ is the jump function
(\ref{eq:jump-def}).  Indeed, the candidate satisfies
(\ref{eq:lXlJK}), provided $s$ has the symmetries of the function $b$.
In the very case of $\sgr{X}=0$ the flow of $\ellJK$ preserves (see
\eqref{eq:symm-JK-comm}) $X$ so the 'jump' function $s$ is constant on
each null curve.  On the other hand, the  preserving of $\ellJK$ by
the flow of $X$, implies then that $s$ is constant along the orbits.
Hence,
\begin{equation}
  [\ell_{(X)},X]\ = 0 \ .
\end{equation}

To show that the orbits of $X-\ell_{(X)}$ are diffeomorphic to $S^1$
we develop the following construction.

Consider a non-degenerate orbit $\hat{\gamma}$ of the isometry
group generated in $\bas$ by the vector field $\hat{X}$ $(=\spi_*X)$.
Let
\begin{equation}
  C_{(\gamma)}\ =\ \spi^{-1} (\hat{\gamma}) \ ,
\end{equation}
that is $ C_{(\gamma)}$ is the cylinder formed by
all the null curves in $\ih$ which correspond to points
of $\hat{\gamma}$.
If $s=0$ on $C_{(\gamma)}$ then the orbits of $X$  are closed
and on the other hand $\ell_{(X)}=0$ so $X-\ell_{(X)}=X$.
Therefore suppose
\begin{equation}
  s\ \neq\ 0\quad \text{ on }\ C_{(\gamma)} \ .
\end{equation}

The idea is to construct a function $v: C_{(\gamma)}\rightarrow \re$
such that
\begin{equation}\label{eq:XvLv}
  \ell_{(X)}^aD_a v\ =\ \frac{s}{2\pi}\ =\  X^aD_a v \ .
\end{equation}
That is sufficient condition for the orbits of
$X$ contained in the cylinder to be closed.

We construct $v$ as follows. On a single null curve
$c_o\subset C_{(\gamma)}$ define $v$ to be
\begin{equation}
  v{|_{c_0}}\ :=\ \bar{v} \ ,
\end{equation}
where $\bar{v}$ is the coordinate compatible with $\ellJK$.
On the curve
\begin{equation}
  c_\phi\ =\ U_\phi(c_{0})
\end{equation}
for $0<\phi<2\pi$ define $v$ by the pullback from $c_0$ {\it plus}
$\frac{s}{2\pi}\phi$, namely
\begin{equation}
  v{|_{c_\phi}}\ =\ (U_{-\phi})^*\left( v{|_{c_0}}\right)
    + \frac{s}{2\pi}\phi \ .
\end{equation}
Due to the definition of the jump function $s$, the resulting
function $v$ is differentiable at every point of the curve
$c_{2\pi}$ as many times as $\bar{v}|_{C_{(\gamma)}}$.

It is also easy to see that $v$ satisfies \eqref{eq:XvLv}.

\subsubsection{Non-extremal case}

In general the diffeomorphism flow $U_{\phi}$ generated by considered
infinitesimal symmetry transforms the coordinate $\bar{v}$ compatible
with J-K null field as indicated by \eqref{eq:JK-u-tr}. This and the
condition $\sgr{X}\neq 0$ imply that there exists exactly one slice
$\slc_o$ of $\ih$ invariant with respect to action of $U_{2\pi}$, that
is
\begin{equation}
  {\exists !}_{\slc_o\subset\ih}\quad \forall_{p\in\slc_o}\quad
  U_{2\pi}^*\bar{v}(p)\ =\ \bar{v}(p) \ .
\end{equation}
The coordinate $\bar{v}$ takes on points $p\in\slc_o$ the following
values:
\begin{equation}
  \forall_{p\in\slc_o}\quad \bar{v}(p)\
  =\ \frac{-\spi^*\hat{b}(\spi(p))}{e^{2\pi\sgr{X}}-1} \ ,
\end{equation}
where $b(p)=:\spi^*\hat{b}(\spi(p))$ (with $b$ being the function
defined via \eqref{eq:JK-u-tr}). One can then easily construct another
coordinate $v$ compatible with $\ellJK$ and such that $\slc_o =
\{p\in\ih:\ v(p)=0 \}$. It is indeed given by the formula
\begin{equation}\label{eq:hel-new-v}
  v(p)\ :=\ \bar{v}(p)
    + \frac{\spi^*\hat{b}(\spi(p))}{e^{2\pi\sgr{X}}-1} \ .
\end{equation}
The diffeomorphism $U_{\phi}$ transforms the new coordinate as follows
\begin{equation}\label{eq:slc-0-pres}
  U(\phi)v\ =\ e^{\sgr{X}\phi} v \ .
\end{equation}

Let us now choose the field $\ell_{(X)}$ such that
\begin{equation}
  \ell_{(X)}\ :=\ \sgr{X}v\ellJK \ .
\end{equation}
Due to \eqref{eq:symm-JK-comm} the field $X-\ell_{(X)}$ commutes with
$\ellJK$
\begin{equation}\label{eq:l0-Phi-comm}
  [\ellJK,X-\ell_{(X)}]\ =\ 0 \ ,
\end{equation}
so the flow $[\ellJK]$ maps each orbit of $X-\ell_{(X)}$ onto
another one. Also on the slice $\slc_o$ the considered field
($X-\ell_{(X)}$) is tangent to it (as action $U_\phi$ preserves the
slice due to \eqref{eq:slc-0-pres}). That, together with the fact that
the flow $[\ellJK]$ maps $\slc_o$ onto $\slc_{v}:=\{p\in\ih:\
v(p)=\const\}$ imply that $X-\ell_{(X)}$ is tangent to each surface
$v=\const$. Its orbits are then closed and diffeomorphic to $S(1)$
(thus $\ell_{(X)}$ satisfies the property \eqref{it:Hawk-orb} of
definition \ref{def:hawking}). Also the flow of $\ell_{(X)}$ preserves
the infinitesimal symmetry $X$
\begin{equation}
  [\ell_{(X)},X]^a\ =\ [\ell_{(X)},X-\ell_{(X)}]^a\
  =\ v[\ellJK,X-\ell_{(X)}]^a - (X^b-\ell_{(X)}^b)D_bv\ellJK^a
  \ =\ 0 \ ,
\end{equation}
and its surface gravity is constant
\begin{equation}
  \sgr{\ell_{(X)}}\ =\ \sgr{X}\ellJK^aD_av\ =\ \sgr{X} \ ,
\end{equation}
so it finally satisfies all the requirements for a Hawking null vector
(see definition \ref{def:hawking}).

The subsection can be summarized by the following:
\begin{cor}\label{cor:hel-ell-sym}
  Suppose a NEH $\ih$ admits an infinitesimal helical symmetry $X$.
  Then on $\ih$ there exists a Hawking null vector field, that is the
  null field $\ell_{(X)}$ satisfying the requirements of definition
  \ref{def:hawking}. Considered field is unique.
\end{cor}
\begin{proof}[Proof of the uniqueness]
Suppose the fields $\ell_{(X)}$, $\ell_{(X)}'$ satisfy definition
\ref{def:hawking}. Then the equations (\ref{eq:lXlJK},
\ref{eq:sgr-ell-X}) imply that these fields are of the form:
\begin{subequations}\label{eq:lXlX'}\begin{align}
  \ell_{(X)}\ &=\ (\sgr{\ell_{(X)}} \bar{v} + b)\ellJK \ , &
  \ell_{(X)}'\ &=\ (\sgr{\ell_{(X)}} \bar{v} + b')\ellJK \ , &
  \lie_{\ellJK}b\ &=\ \lie_{\ellJK}b'\ =\ 0 \ .
    \tag{\ref{eq:lXlX'}}
\end{align}\end{subequations}
The condition \ref{eq:Hawk-comm} imposes the following relation
between $b, b'$
\begin{equation}
  X^aD_a(b-b')\ =\ 0 \ ,
\end{equation}
so $b-b'$ is constant along cylinders $\ih\supset C_{(\gamma)} :=
\spi^{-1}(\hat{\gamma})$ built over nontrivial orbits $\hat{\gamma}$
of infinitesimal symmetry $\hat{X}$ induced on $\bas$.

Using the same method (of averaging over a diffeomorphism group
generated via \eqref{eq:symm-diff} by a vector field) as
the one used in the proof of lemma \ref{lem:l0-phi-comm} one can show,
that there exists a coordinate system $\bar{v}$ compatible with
$\ellJK$ such that orbits of $X-\ell_{(X)}$ (closed due to property
\eqref{it:Hawk-orb} of definition \ref{def:hawking}) lie on constancy
surfaces of $\bar{v}$. One can then immediately generalize corollary
\ref{cor:axi-form} to the case of $\Phi:=X-\ell_{(X)}$ (where $\Phi$
is not necessarily an infinitesimal symmetry). Thus there exists the
diffeomorphism $h:\ih\to\bas\times\re$ such that
\begin{equation}
  h_{*}(X-\ell_{(X)}')\ =\ (\hat{X},(b'-b)\partial_{\bar{v}}) \ .
\end{equation}
It implies immediately, that the field $X-\ell_{(X)}'$ can
satisfy the property \ref{it:Hawk-orb} of definition \ref{def:hawking}
if and only if $b = b'$. That statement completes the proof.
\end{proof}

In the next subsection we will show that the Hawking vector
constructed above is in fact null symmetry at $\ih$.

\subsection{Induced null symmetry}
  \label{sec:hel-ind-null-symm}

In this subsection we assume that the matter energy-momentum tensor
$T_{ab}$ satisfies the condition \eqref{eq:mat-cond} for some,
non-vanishing null $\ell\in\Gamma(T(\ih))$.

We will show that the Hawking vector field constructed in the previous
subsections is an infinitesimal symmetry. The about $T_{ab}$ implies
immediately that whenever the field $\ell_{(X)}$ (defined in corollary
\ref{cor:hel-ell-sym}) doesn't vanish, its flow as well as the flow of
$X$ preserve the rotation $1$-form corresponding to $\ell_{(X)}$
\begin{equation}\label{eq:hel-w-symm}
  \lie_{\ell_{(X)}}\w{\ell_{(X)}}\ =\ - \lie_{X}\w{\ell_{(X)}}\
    =\ 0 \ .
\end{equation}

Denote the set of points of $\ih$ at which $\ell_{(X)}\neq 0$ by
$\mathcal{U}$. Due to \eqref{eq:Hawk-comm} the field $\ell_{(X)}$
admits at $\mathcal{U}$ a coordinate $v'$ compatible with $\ell_{(X)}$
and constant along the orbits of $\Phi_{(X)}$ so $X$ preserves
foliation of $\mathcal{U}$ by surfaces $v'=\const$.\footnote{In the
  systematic development of the Hawking null field presented in Appendix
  \ref{ssec:Hawk-syst} that property is just part of the definition.}

Given this coordinate one can define a transversal covector field
$n'_a=-D_a v'$. It is (again due to \eqref{eq:Hawk-comm})
preserved by the considered symmetry. In consequence the transversal
expansion-shear tensor $\tS_{AB}$ corresponding to $n'_a$
(orthogonal to the surfaces $v'=\const$) is also preserved by the flow
of $X$
\begin{equation}\label{eq:X-tS}
  \lie_{X}\tS_{AB}\ =\ 0 \ .
\end{equation}
On the other hand given a null field of a constant surface gravity the
evolution (along null geodesics) of $\tS_{AB}$ corresponding to a
coordinate compatible with it is determined by the equation
\eqref{eq:transexpsh}. In the case considered here that equation
determines the evolution of $\tS_{AB}$ corresponding to $v'$
\begin{equation}\label{eq:phi-tS}
  \tS_{AB}\ =\ \begin{cases}
    \begin{split}
      &v'\, \tq^{a}{}_{A}\tq^{b}{}_{B}
        \left( (\spi^*\hD\hw{\ell_{(X)}})_{(ab)}
               + (\spi^*\hw{\ell_{(X)}})_a(\spi^*\hw{\ell_{(X)}})_a
    \right)+ \\
      &\quad +\, v'\ \left( - \fracs{1}{2}\,\Ricn{n-2}_{AB}
                        - \fracs{1}{2}\Lambda \tq_{AB}
                 \right) \\
      &\quad +\ \tq^{a}{}_{A}\tq^{b}{}_{B}
        \left( (\spi^*\hS^o)_{ab}
                   + \fracs{1}{2}(\spi^*\hat{T})_{ab} \right)
    \end{split} \ , & \sgr{\ell_{(X)}}=0 \ , \\
    \begin{split}
      &\frac{1}{\sgr{\ell_{(X)}}}\tq^{a}{}_{A}\tq^{b}{}_{B}
        \left((\spi^*\hD\hw{\ell_{(X)}})_{(ab)}
          + (\spi^*\hw{\ell_{(X)}})_a(\spi^*\hw{\ell_{(X)}})_b
    \right)+  \\
      &\quad +\, \frac{1}{\sgr{\ell_{(X)}}}\left(
        - \fracs{1}{2}\,\Ricn{n-2}{}_{AB}
        - \fracs{1}{2}\Lambda \tq_{AB}
    + \fracs{1}{2}\tq^{a}{}_{A}\tq^{b}{}_{B}(\spi^*\hat{T})_{ab}
    \right)\\
      &\quad +\ e^{-\sgr{\ell_{(X)}} v'}
            \tq^{a}{}_{A}\tq^{b}{}_{B}(\spi^*\hS^o)_{ab}
    \end{split} \ , & \sgr{\ell_{(X)}}\neq 0 \ ,\\
  \end{cases}
\end{equation}
where all the objects are defined analogously to the ones used
in \eqref{eq:transexpsh}.

The equation \eqref{eq:X-tS} implies that $U_{2\pi}^*\tS_{AB} =
\tS_{AB}$ so the term proportional to $v'$ (for $\sgr{X}=0$) or the
component $\spi^*\hS^o{}_{AB}$ (otherwise) vanish respectively. Thus
\begin{equation}\label{eq:chir-axi-ih}
  \lie_{\ell_{(X)}}\tS_{AB}|_{\mathcal{U}}\ =\ 0 \ .
\end{equation}

We have established,
\begin{equation}\label{eq:desire}
  [\ell_{(X)}, D]\ =\ 0
\end{equation}
on the subset $\mathcal{U}\subset\ih$ such that $\ell_{(X)}\neq 0$.
Therefore \eqref{eq:desire} holds also on the closure
$\bar{\mathcal{U}}$. Consider then the set $\ih\setminus
\bar{\mathcal{U}}$. Since this set is open in $\ih$ and
\begin{equation}
  \ell_{(X)}\ =\ 0 \ ,
\end{equation}
in it, it follows that \eqref{eq:desire} holds in $\ih\setminus
\bar{\mathcal{U}}$ as well.

Finally both the fields $\ell_{(X)}$ and $X-\ell_{(X)}$
are infinitesimal symmetries at $\ih$:
\begin{theorem}\label{thm:hel-decomp}
  Suppose a non-expanding horizon $\ih$ is equipped with a
  energy-momentum tensor $T_{ab}$ such that the condition
  \eqref{eq:mat-cond} holds for arbitrary null field $\ell$
  tangent to the horizon.
  If considered $\ih$ admits a helical infinitesimal symmetry $X$,
  then it also admits a cyclic infinitesimal symmetry $\Phi$
  and a null infinitesimal symmetry $\bsl$ such that
  \begin{equation}
    X\ =\ \Phi + \bsl \ .
  \end{equation}
\end{theorem}

The existence of null symmetry implies immediately via theorems
\ref{thm:ell-nex} and \ref{thm:ell-ex} that the set of points
intersected by open orbits of $X$ is dense at $\ih$.

\section{Symmetric NEHs in 4D spacetime}
  \label{sec:neh4d}

In the studies carried out through sections \ref{sec:symm-gen-null}
till \ref{sec:symm-gen-chir} nothing was assumed about the horizon
dimension or topology of its base space. Also assumed energy
conditions allowed quite broad variety of matter fields. In this
section we will restrict our studies to NEHs embedded in $4$-dimensional
spacetime and whose base space is diffeomorphic $S^2$. In most cases
we will restrict possible matter fields to Maxwell field only
(including the zero electromagnetic field) also assuming then, that
the cosmological constant vanishes. In these cases he notion of a
symmetry will be strengthened: Besides the properties enlisted in
definition \ref{def:symm} we will require the preservation of an
electromagnetic field tensor $F$, namely:
\begin{equation}\label{eq:lielF}
  \lie_X \cF_a{}^\mu \ =\ 0 \ ,
\end{equation}
where $\cF:=F-i\star F$ and $X$ is an infinitesimal symmetry.

In the case described above a complete classification of the possible
infinitesimal symmetries will be derived.

The description of the geometry of an electrovac NEH in
$4$-dimensional spacetime is briefly presented in Appendix
\ref{sec:evac-geom}. For the computational convenience we use
the Newman-Penrose tetrad formalism in both the geometry
description and the analysis of symmetric electrovac NEHs presented
here. Thus in the present section the index notation will be
dropped.

As Maxwell field satisfies condition \eqref{eq:mat-cond} the results
established in sections \ref{sec:symm-gen-null} through
\ref{sec:symm-gen-chir} hold in particular for symmetries considered
here without additional energy assumptions. Moreover (which will be
shown in subsection \ref{sec:EV-hel}) the infinitesimal symmetries
induced on the horizon by helical one (see theorem
\ref{thm:hel-decomp}) satisfy \eqref{eq:lielF} (provided the helical
infinitesimal symmetry satisfies that condition).

The stronger definition of the symmetry allows us also to improve
the general results in two cases: extremal null symmetry (theorem
\ref{thm:ell-ex}) and NEHs admitting $2$-dimensional null symmetry
group (section \ref{sec:nullsym-general}). We will focus on the
former case first.

\subsection{Extremal null symmetry of an electrovac NEH in 4D}
  \label{sec:nullsym-ex-evac}

In this subsection we consider electrovac NEHs in $4$-dimensional
spacetime, however the requirement for the horizon base space to be
diffeomorphic with $2$-sphere is relaxed.

\begin{theorem}\label{thm:ell-ex-evac}
  Suppose $\ih$ is a (maximal analytic extension of a) non-expanding
  horizon embedded in a $4$-dimensional spacetime satisfying the
  Einstein-Maxwell (including vacuum Einstein) field equations with
  vanishing cosmological constant. Suppose also $\bsl$ is an extremal
  null infinitesimal symmetry in the sense of definition \ref{def:symm}
  strengthened by condition \eqref{eq:lielF}. Then $\bsl$ vanishes
  nowhere at $\ih$.
\end{theorem}
\begin{proof}
Any extremal null symmetry can be expressed in terms of
Jezierski-Kijowski vector field via \eqref{eq:ex-symm-JK}.
On the other hand the condition (\ref{eq:symm_def}b) implies via
(\ref{eq:[l,D]1}, \ref{eq:com_form}, \ref{eq:mulS}, \ref{Dmu},
\ref{eq:cevRic}b) the following constraint
\begin{equation}\label{eq:ex-constr}
    \fracs{1}{2}\Ricn{2}\ :=\ \fracs{1}{2}\hq^{AB}\Ricn{2}_{AB}\
      =\ 2|\Phi_1|^2 \
      +\ \tdiv(\hw{\ellJK} + \td\ln \hat{B})
      + |\hw{\ellJK} + \td\ln B|^2_{\tq}\ ,
\end{equation}
true everywhere where $B$ doesn't vanish. The $0$th Law allows us to
re-express this equation as a constraint defined at $\bas$ and
involving projective data just by replacing projected objects by
projective ones. As (due to theorem \ref{thm:ell-ex}) $B\neq 0$ at
dense subset of $\bas$ the re-expressed constraint can be integrated
over $\bas$. Also the integral $\int \Ricn{2}\hvol$ is determined by
Gauss-Bonnet theorem. Therefore the equation \eqref{eq:ex-constr}
takes form of topological constraint:
\begin{equation}\label{eq:top-cen}
  4\pi(1-\boldsymbol{g})\
  =\ \int_{\bas} \left( |\hw{\ellJK} + \hd\ln B|^2_{\hq}
    + 2|\Phi_1|^2 \right)\hvol \ ,
\end{equation}
where $\boldsymbol{g}$ is the genus of $\bas$.

Finally the only topologies of the base space allowed in this case are
$S^2$ and $S^1\times S^1$. Moreover for $\bas=S^1\times S^1$ the only
allowed solution is $(\hRicn{2}_{AB} = 0,\ \hw{\bsl} = 0,\ \hcF_{AB} =
0)$ (where $\hcF$ is defined via \eqref{eq:hcF}). The vanishing of
$\hw{\bsl}$ (in particular its exact part) implies immediately that
$\bsl$ is (globally) proportional to Jezierski-Kijowski null field so
it nowhere vanishes.

In the case $\bas=S^2$ the rotation 1-form $\hw{\ellJK}$
(corresponding to J-K null vector field $\ellJK$) is a co-exact
$1$-form at $\bas$ (both the harmonic and exact part vanish
identically). The system \eqref{eq:nullsym-full} can be then written
as
\begin{equation}\label{eq:nullsym-full-ex}\begin{split}
  \left[ \hD^{}_A\hD^{}_B + 2\hw{\ellJK}^{(co)}_{(A}\hD^{}_{B)}
         + (\hD^{}_{(A}\hw{\ellJK}^{(co)}_{B)}) \right] \hat{B}\ &+ \\
  \left[ \hw{\ellJK}^{(co)}_A\hw{\ellJK}^{(co)}_B
         - \fracs{1}{2}\Ricn{n-2}^{}_{AB}
         - \fracs{1}{2}\Lambda\hq^{}_{AB} + \fracs{1}{2}\hat{T}^{}_{AB}
  \right] \hat{B} \ &= \ 0 \ ,
\end{split}\end{equation}
where $\hat{T}_{AB}$ is a projective energy-momentum tensor of the
Maxwell field. In the null frame introduced in appendix
\ref{sec:evac-geom} the trace and traceless part of the above equation
form the following system
\begin{subequations}\label{eq:nsf-frame}\begin{align}
  \label{eq:nsf-frame-trace}
    \left(\hmo\hmob + \hmob\hmo - 2a\hmo - 2\bar{a}\hmob
      + 2i \hmo U \hmob - 2i \hmob U \hmo
      + 2 \hmo U \hmob U - \fracs{1}{2}\Ricn{2} \right) \hat{B}\
      + \ 2|\Phi_1|^2 \hat{B} \
      &=\ 0 \ ,  \\
  \label{eq:nsf-frame-trless}
    \left(\hmob\hmob + 2a\hmob -2i \hmob U\hmob -i\hmob\hmob U
      - 2ia\hmob U - \hmob U\hmob U  \right) \hat{B}
    \ &=\ 0 \ ,
\end{align}\end{subequations}
where $U$ is a rotation potential defined via \eqref{eq:UBa}, $a$ is a
component of the Levi-Civita connection $\hat{\Gamma}$ on $\bas$
defined via \eqref{2gamma} and $\hmo$ is given by
\eqref{eq:delta_def}.

In vacuum case by commuting the $\hmo$ operator with  $\hmo\hmob +
\hmob\hmo$ as well $\hmob$ with $\hmo\hmo$ one gets the following
integrability condition for \eqref{eq:nsf-frame}
\begin{equation} \label{eq:bian}
  \hmob\Psi_2 + 3\bpi_o\Psi_2\ =\ 0
\end{equation}
where $\bpi_o$ is a coefficient (defined by \eqref{eq:w-frame}) of
$\hw{\ellJK}$ in decomposition with respect to used frame and $\Psi_2$
is an invariant complex scalar 
\begin{equation}\label{eq:Psi2-loc-def}
  \Psi_2\ =\ \frac{1}{2}\left( -\fracs{1}{2}\Ricn{2}
  + i\hat{\Delta}U \right) \ ,
\end{equation}
(with $\hat{\Delta}$ being a Laplace operator on $\bas$).

On the other hand preserving the electromagnetic field tensor
(expressed now in terms of coefficients $\Phi_I$ defined via
\eqref{F}) by $\bsl=(\spi^*\hat{B})\ellJK$ \eqref{eq:lielF} imposes via
one of Maxwell field equations \eqref{eq:DPhi2} the
constraint \eqref{eq:ex-M-c} which may be written as a pull-back from
$\bas$ of
\begin{equation}\label{eq:nullsym-Max-ex}
  \hmob\hat{\Phi}_1 + 2\bpi_o\hat{\Phi}_1
    + 2(\hmob\ln\hat{B})\hat{\Phi}_1
  \ =\ 0\ ,
\end{equation}
where the component $\bpi$ corresponding (via
eq. \eqref{eq:w-frame}) to $\hw{\bsl}$ (well defined on a dense subset
of $\bas$ via Theorem \ref{thm:ell-ex}) was expressed in terms of
$\bpi_o$.\footnote{The transformation $\hw{\ell}_A\to\hw{\ell}_A+D_Af$
  in terms of frame component $\bpi$ takes the form
  $\bpi\to\bpi+\hmob f$.}

The Hodge decomposition of $\hw{\ellJK}$ can be written as
decomposition (\ref{eq:UBa}b) of $\bpi_o$. Applied to equations
(\ref{eq:bian}, \ref{eq:nullsym-Max-ex}) it allows their explicit
integration which produces the following constraints on NEH
invariants and function $\hat{B}$ valid in vacuum
(\ref{eq:nullsym-const}a) and electrovac (\ref{eq:nullsym-const}b)
case respectively:
\begin{subequations}\label{eq:nullsym-const}\begin{align}
  \hat{B}^3|\Psi_2| &= C_0 = \const\ , &
  \hat{B}^2|\hat{\Phi}_1| &= E_0 = \const\ .
    \tag{\ref{eq:nullsym-const}}
\end{align}\end{subequations}
As the coefficients $\Psi_2$ and $\hat{\Phi}_1$ are finite we need
only to show that the constants $C_0$, $E_0$ for appropriate case are
non-zero.

Let us start with vacuum case first. As the scalar $\Psi_2$ cannot
identically vanish on the entire $\bas$ the identity $C_0=0$ implies
that there exists a closed subset of $\bas$ (with non-empty interior)
on which $\hat{B}=0$. It is however excluded by corollary
\ref{thm:ell-ex}.

In electrovac case on the other hand there exists an open subset of
$\bas$ on which $\hat{\Phi}_1\neq 0$ (otherwise $\hat{T}_{AB}=0$ in
\eqref{eq:nullsym-full-ex} and we end up with just a vacuum
constraint). The condition $E_0=0$ requires then vanishing of
$\hat{B}$ on this subset. It is again excluded by corollary
\ref{thm:ell-ex}.

Finally an appropriate for considered case (vacuum or electrovac)
constant in \eqref{eq:nullsym-const} can take only non-zero
value. That fact together with the finiteness of the coefficients
involved in \eqref{eq:nullsym-const} ensures that $\hat{B}$ nowhere
vanishes.
\end{proof}

it was shown in \cite{ex} that the maximal group of extremal null
symmetries on a given NEH is exactly $1$-dimensional provided the
symmetries are generated by nowhere vanishing infinitesimal
symmetries. The theorem \ref{thm:ell-ex-evac} allows us to relax the
last condition:
\begin{prop}\label{prop:ExUniqEvac}
  Suppose $\bsl$, $\bsl'$ are extremal null infinitesimal symmetries
  (satisfying also \eqref{eq:lielF}) of an electrovac NEH $\ih$ in
  $4$-dimensional spacetime (with cosmological constant
  $\Lambda=0$). Then
  \begin{equation}
    \bsl\ =\ c\bsl'\ .
  \end{equation}
  where $c\ =\ \const$.
\end{prop}

As the geometry of every NEH satisfying the assumptions of theorem
\ref{thm:ell-ex-evac} has to satisfy the constraints
(\ref{eq:nsf-frame},\ref{eq:nullsym-Max-ex}) the set of possible
solutions is seriously restricted. In particular all the axial
solutions are given by the projective metric, rotation $1$-form and
projective electromagnetic field tensor of the extremal horizon of
Kerr-Newman metric. On the other hand it was shown in \cite{crt} that
the only vacuum extremal IHs in four dimensions whose rotation
$2$-form vanishes (denoted as {\it non-rotating}) are trivial
solutions of a toroidal base space and $(\Ricn{2}=0,\ \hw{\bsl}_A=0,\
\hat{\Phi}_1=0)$. Indeed,
due to (\ref{eq:top-cen}) the only possible topologies of the horizon
base space are $2$-sphere and $2$-torus (with only trivial solution
allowed in the latter case). In the case of $\bas=S^2$ $\hat{B}$ and
$\Ricn{2}$ are constrained by (\ref{eq:nullsym-const}a) which in
non-rotating case takes the form\footnote{In general case (rotation)
  an integration of \eqref{eq:bian} gave complex expression involving
  $\Psi_2$, $\hat{B}$ and $U$. The equation (\ref{eq:nullsym-const}a)
  was obtained by taking the absolute value of the result of
  integration. In the non-rotating case the result of integration is
  real up to constant phase which can be fixed by gauge transformation
  $U\to U+U_0$, $U_0=\const$. Thus instead of taking the absolute
  value we use the integration result itself.}
\begin{equation}\label{eq:ns-const-nrot}
  \Ricn{2}\hat{B}^3\ =\ 4C_0\ =\ \const \ ,
\end{equation}
where $C_1\neq 0$. As $\hat{\Omega}_{AB}=0\ \Rightarrow\ U=\const$
the substitution of $\Ricn{2}$ in \eqref{eq:nsf-frame-trace} by
\eqref{eq:ns-const-nrot} gives us the following elliptic PDE
\begin{equation}
  \left[\hat{\Delta} - \frac{C_0}{\hat{B}^3}\right]\hat{B}\ =\ 0 \ ,
\end{equation}
where $\hat{\Delta}$ is the Laplasian on the base space. An
integral over $\bas$ of the above PDE
\begin{equation}
  \int_{\bas}\hvol\left[\hat{\Delta}
    - \frac{4C_0}{\hat{B}^3}\right]\hat{B}\
  =\ -4\int_{\bas}\frac{C_0}{\hat{B}^2}\hvol\ =\ 0 \ ,
\end{equation}
implies then $C_0=0$. This case was however excluded (see discussion
after \eqref{eq:nullsym-const}). Thus the following is true:
\begin{cor}\label{cor:nrex-sol}
  Suppose $\ih$ is an electrovac NEH embedded in $4$-dimensional
  spacetime (with $\Lambda=0$). Suppose also it admits an extremal
  null infinitesimal symmetry $\bsl_o$ and its rotation $2$-form
  vanishes. Then its geometry is given by the data
  \begin{subequations}\label{eq:nrex-sol}\begin{align}
    \Ricn{2}\ &=\ 0 \ , & \hw{\bsl_o}\ &=\ 0 \ , &
    \hat{\Phi}_1\ &=\ 0 \ , \tag{\ref{eq:nrex-sol}}
  \end{align}\end{subequations}
  whereas the base space of $\ih$ is a $2$-torus.
\end{cor}

\subsection{Electrovac NEHs admitting $2$-dimensional null symmetry
  group}
  \label{sec:EV-2null}

The class of electrovac isolated horizons admitting $2$-dimensional
group of null symmetries was investigated in \cite{ex}. Theorem
\ref{thm:ell-ex-evac} allows us to directly apply the results
presented there to more general case considered in this article.
Indeed the following is true:

\begin{prop}\label{thm:ExNull--}
  Suppose an electrovac NEH in $4$-dimensional spacetime (with
  $\Lambda=0$) $\ih$ admits a two-dimensional group of null
  symmetries (generated by vector fields satisfying definition
  \ref{def:symm} and \eqref{eq:lielF}). Then $\ih$ is an extremal
  Isolated Horizon (i.e. there exists an extremal nowhere vanishing
  infinitesimal null symmetry satisfying \eqref{eq:lielF}).
\end{prop}

The statement above implies automatically, that the base space of
considered horizon is either $2$-sphere or $2$-torus, where the latter
case contains only trivial solution \eqref{eq:nrex-sol}.

\begin{prop}
  A general nontrivial electrovac (including vacuum) extremal IH $\ih$
  (with an infinitesimal symmetry $\bsl_o$) in $4$-dimensional
  spacetime (with $\Lambda=0$) admitting an additional null symmetry
  is given by any solution to (\ref{eq:nsf-frame},
  \ref{eq:nullsym-Max-ex}), $\mu=0=\lambda$ in \eqref{G32} and
  $\Phi_2=0$ \footnote{This condition wasn't present in an analogous
    theorem in \cite{ex} because only one of two distinct
    infinitesimal symmetries was required to preserve $\cF_a{}^{\mu}$
    there. The proof that $\Phi_2=0$ if all the infinitesimal
    symmetries preserve $\cF_a{}^{\mu}$ is analogous to proof of
    constraint $\mu=0=\lambda$.
  }.
  Its group of the null symmetries is exactly two-dimensional, the
  generators are an infinitesimal symmetry $\bsl_o$ and $\bsl=v\bsl_o$
  where $v$ is a coordinate compatible to $\bsl_o$ and such that the
  transversal expansion-shear tensor corresponding to it
  vanishes\footnote{This condition is equivalent to $\mu=0=\lambda$,
    see appendix \ref{sec:evac-geom}} at $\ih$.
  The commutator between the generators is
  \begin{equation}
    [\bsl_o,\bsl]\ =\ \bsl_o.
  \end{equation}
\end{prop}

\subsection{Helical electrovac NEHs}
  \label{sec:EV-hel}

The quasi-local rigidity theorem \ref{thm:hel-decomp} developed for the
general NEH holds in particular for electrovac horizons in
$4$-dimensional spacetime. Moreover the induced infinitesimal
symmetries satisfy the condition \eqref{eq:lielF} (provided it is
satisfied by the helical symmetry). To show that it is enough to check
the constancy of the component $\Phi_2$\footnote{The other components
  remain constant along null geodesics at $\ih$.} of $\cF_{a}{}^{\mu}$
defined by \eqref{F} with respect to the frame $(e_1,\ldots,e_4)$
defined in \ref{sec:frame-H} chosen such that $e_4^{\mu}
=\ell_{(X)}^{\mu}$ \footnote{$\ell_{(X)}^{\mu}$ is a completion of
  $\ell_{(X)}^a$ to null vector in $T(\M)$ at $\ih$. Due to theorems
  \ref{thm:ell-nex} and \ref{thm:ell-ex} considered frame is defined
  on dense subset of $\ih$.}.
Indeed due to equation \eqref{eq:DPhi2} $\Phi_2$ is either exponential
($\sgr{\ell_{(X)}}\neq 0$) or polynomial (otherwise) in coordinate
compatible with $\ell_{(X)}$, thus it is constant along the horizon
null geodesics by an argument similar to the one used in proof of
constancy of $\tS_{AB}$ in subsection \ref{sec:hel-ind-null-symm}. The
following is then true:
\begin{theorem}\label{thm:EV-rigid}
  Suppose an electrovac NEH $\ih$ embedded in a $4$-dimensional
  spacetime (with $\Lambda=0$) admits a helical symmetry. Then it
  admits also a null and axial symmetry which also preserve the
  electromagnetic field tensor $F_a{}^{\mu}$. Thus, depending on the
  value of $\sgr{X}$, either:
  \begin{enumerate}[ a)]
    \litem the entire $\eih$ in the case $\sgr{X}=0$, or
    \litem each of two sectors $\ih_{\pm}=\{p\in\ih:\ \sgn(v(p))=\pm
      1\}$ (where $v$ is defined via \eqref{eq:hel-new-v})
      in the case $\sgr{X}\neq 0$
  \end{enumerate}
  constitutes an axial isolated horizon\footnote{I.e. the null
    symmetry vanishes nowhere at it.}.
\end{theorem}

\subsection{Classification of the symmetric horizons in $4$-dimensional
  spacetime}
  \label{sec:4D-class}

The general properties of distinguished classes of symmetric NEH
investigated in the previous part of this article allow us to
introduce the complete classification of symmetric non-expanding
horizons embedded in a $4$-dimensional space-time, provided  the
base space  topology is $S^2$.\footnote{A classification of
  symmetric, non-extremal weakly isolated horizons was developed in
  \cite{abl-g,cz}. However, the symmetries considered therein were
  defined as preserving the induced metric $q_{ab}$ and the null flow
  $[\ell]$ of a given non-extremal weakly isolated horizon. That
  difference is essential and simplifies the classification
  considerably.
}

Given a NEH $\ih$ whose base space $\bas$ is diffeomorphic to a
$2$-sphere, any vector field $X^a$ which generates a symmetry of
$\ih$ induces on $\bas$ a Killing vector field $\hat{X}^A$ of the
following properties: 
\begin{itemize}
  \item $\hat{X}^A$ is of the form $\hat{X}^A=\hvol^{AB}\hD_Bh$, where
    $h$ is a function defined on $\bas$.
  \item Since all the 2-sphere metrics are conformal to the round
    metric on that sphere, the $\hat{X}^A$ is a conformal Killing
    field of the round $2$-sphere metric.
\end{itemize}
Applying the classification of conformal Killing fields on the
2-sphere to the Killing fields on $\hat{\ih}$ we can divide them 
onto the following classes: 
\begin{enumerate}[ (i)]
  \litem a rotation, \label{it:C-rot}
  \litem a boost, \label{it:C-boo}
  \litem a null rotation (with exactly one critical point of
    $[\hat{X}]$), \label{it:C-null}
  \litem a linear combination of the representatives of the classes
    \eqref{it:C-rot}-\eqref{it:C-null}. \label{it:C-lc}
\end{enumerate}
In the cases \eqref{it:C-boo},\eqref{it:C-null} and \eqref{it:C-lc}
all the orbits of $\hat{X}$ converge to one critical point. The
function $h$ is then constant on entire $\bas$, hence in all the cases
except \eqref{it:C-rot} the projected symmetry necessarily vanishes.
Finally the metric of the base space necessarily belongs to one of the
following classes:
\begin{enumerate}[ (1)]
  \item spherical: group of symmetries is 3-dimensional
    group of rotations \label{it:CH-s}
  \item axial: 1-dimensional group of rotational
    symmetries
  \item generic: $0$-dimensional group of symmetries \label{it:CH-g}
\end{enumerate}
The above statements imply that the field $\hat{X}$ induced on $\bas$
by a symmetry $X$ either is the rotational Killing field or
identically zero. Thus given a symmetric NEH all of symmetries it
admits necessarily belong to one of the following classes:
\begin{itemize}
  \litem Null symmetries (see definition \ref{def:null-def}): Horizons
    admitting null symmetry generated by nowhere vanishing vector
    field are referred to as isolated horizons and were discussed
    extensively in the literature \cite{abdfklw,abl-g}
    The general null-symmetric NEHs were discussed in section
    \ref{sec:symm-gen-null}. Note that a given NEH can admit
    the $2$ dimensional non-commutative group of null symmetries.
    This case was discussed (in context of an IH geometry) in
    \cite{ex}.
  \litem Axial symmetries (see definition
    \ref{def:axi-def})\footnote{As $\bas = S^2$ all the cyclic
      symmetries are axial ones}: NEHs admitting such symmetry were
    discussed in section \ref{sec:symm-gen-axi}.
  \litem Helical symmetries (see definition \ref{def:hel-def}):
    Due to theorem \ref{thm:hel-decomp} a helical symmetry induces on
    a NEH both axial and null symmetry.
\end{itemize}
The properties of possible NEH symmetries enlisted above allow us to
introduce a classification complementary to \eqref{it:CH-s} -
\eqref{it:CH-g}, namely we divide symmetric NEHs (with respect to
structure of their groups of null symmetries) onto the following
classes:
\begin{enumerate}[ (a)]
  \item 'Null-multisymmetric' NEH's: the group of null symmetries
    is at least 2-dimensional.\label{it:CX-n}
  \item Null-symmetric NEH's: the null symmetry is unique up to
    rescaling by a constant.
  \item Generic NEH's: without null symmetries.\label{it:CX-g}
\end{enumerate}
All the combinations of \eqref{it:CH-s}-\eqref{it:CH-g} and
\eqref{it:CX-n}-\eqref{it:CX-g} are possible thus allowing to
introduce the complete classification as follows:
\begin{cor}\label{cor:class}
  Suppose $\ih$ is a NEH embedded in $4$-dimensional spacetime
  satisfying Einstein field equations (possibly with a cosmological
  constant and matter such that \eqref{eq:mat-cond} holds). Suppose
  also the base space of $\ih$ is a $2$-sphere. Then $\ih$ necessarily
  belongs to one of classes labeled by two non-negative integers
  $(a,n)$: dimensions of the maximal group of axial and null
  symmetries respectively. This pair uniquely characterizes the
  structure of maximal symmetry group of each horizon.
\end{cor}

In the class of electrovac NEHs (with vanishing cosmological constant)
the group of null symmetries is at most $2$-dimensional
\cite{ex}\footnote{In \cite{ex} only symmetric IHs were considered,
  however due to theorems \ref{thm:ell-nex}, \ref{thm:ell-ex},
  \ref{thm:ell-ex-evac} all the electrovac null-symmetric NEHs in 4D
  are isolated horizons.}, so the case \eqref{it:CX-n} consists of the
NEHs with exactly $2$-dimensional null symmetry group. In this case
the null infinitesimal symmetries $\bsl$ and $\bsl_o$ can be chosen
such that (see also \cite{ex} for details) 
\begin{equation}
  [\bsl,\bsl_o] = \bsl_o \ ,
\end{equation}
where $\bsl_o$ is a unique extremal infinitesimal symmetry.

On the other hand theorem \ref{thm:ell-ex-evac} ensures that any
null symmetry vanishes only at the cross-over surface (when
non-extremal) or nowhere (otherwise).

\section{Summary}

The definition and general properties of the infinitesimal
symmetries of symmetric NEHs were studied in Section
\ref{sec:symm-basic}. Most of the results rely on Einstein's 
equations with a possible cosmological constant and matter
satisfying suitable energy inequalities \eqref{eq:T_cau} and
energy-momentum equalities \eqref{eq:mat-cond}. The vacuum case always
satisfies our assumptions. In the context of non-vacuum symmetric NEHs,
appropriate symmetry conditions (\eqref{eq:lielF} in electrovac case)
are imposed on the matter as well. 

Using the Jezierski-Kijowski invariant local flow, every
infinitesimal symmetry was assigned a certain constant
(\ref{eq:symm-JK-comm}). If the constant is zero,
the infinitesimal symmetry is called extremal. It is called
non-extremal otherwise. Another useful general result is the
observation of proposition \ref{prop:symm-ext} that every symmetric
NEH in question is a segment of an (abstract, not necessarily
embedded) symmetric NEH whose null curves are complete in any
affine parametrization. Moreover, on that analytic extension
of a given symmetric NEH, the infinitesimal symmetry generates
a group of globally defined symmetry maps. Since that observation,
we consider only the symmetric complete analytic extensions
of NEHs.

The general case of a NEH $\ih$  admitting a null infinitesimal
symmetry (null symmetric NEHs) is considered in Section
\ref{sec:symm-gen-null}.  The possible  zero
points of the infinitesimal symmetries are studied with
special care. In the non-extremal null infinitesimal
symmetry case, the zero set is just a single cross-section of
$\ih$ (see theorem \ref{thm:ell-nex}). In the extremal case,
we were only able to prove in theorem \ref{thm:ell-ex}
that the subset of $\ih$ on which the infinitesimal symmetry
does not vanish is dense in $\ih$.

The case of more than one dimensional null symmetry group is
partially characterized by theorem \ref{thm:nullsym-ndim}. It is
shown, that the symmetry group necessarily contains an extremal null
symmetry. In the consequence, the NEHs of that symmetry can be
labeled in the vacuum case by solutions to the extremal null-symmetric
NEH constraint \eqref{eq:nullsym-full}\footnote{With $\hat{T}_{AB}=0$,
  $\Lambda = 0$.}. In particular for spacetime dimension $n=4$
considered NEHs can be labeled by solutions to extremal IH constraint
\eqref{eq:in-extr}. That observation leads to a complete
characterization of a NEH which admits more then one linearly
independent null infinitesimal symmetries.

If a NEH $\ih$ admits an helical infinitesimal symmetry $X$, then
the NEH geometry $(q,D)$ and $X$ determine a certain null vector
field $\ell_{(X)}$ on $\ih$. We refer to that vector field as the
Hawking vector field, because it is a generalization of the vector
field defined by Hawking-Ellis \cite{he} in their proof of the
rigidity theorem. In fact we do not need to prove the uniqueness of
our Hawking field. We just construct it and then show the property
crucial for our considerations: the Hawking vector field is an
infinitesimal symmetry itself. Moreover, the difference vector
$X-\ell_{(X)}$ is a cyclic infinitesimal symmetry. The exact
statement is contained in theorem \ref{thm:hel-decomp}. In
particular, the assumptions are satisfied automatically in every
Einstein-Maxwell case. The uniqueness of the Hawking vector field
is  shown in corollary \ref{cor:hel-ell-sym}.

The application of our results in the standard case: $n=4$,
topologically spherical cross section of $\ih$ and Einstein-Maxwell
equations satisfied at $\ih$, is individually studied in  Section
\ref{sec:neh4d}. Given a symmetric NEH, the electro-magnetic field
on $\ih$ is assumed  to satisfy symmetry condition \eqref{eq:lielF}. 
In this case, every extremal null infinitesimal symmetry nowhere
vanishes at $\ih$ (theorem \ref{thm:ell-ex-evac}). This result is
sufficient to complete the classification of possible symmetry groups
of NEHs generated by null infinitesimal symmetries. Combined with the
main result concerning the helical symmetry as with our knowledge of
the symmetric Riemannian geometries of a 2-sphere, it provides the
complete classification of the symmetric NEH in this case (Section
\ref{sec:4D-class}).

\section*{Acknowledgments}

We would like to thank Piotr Chru\'sciel and Jan Derezi\'nski for
discussions. This work was supported in part by the NSF grants
PHY-0354932 and PHY-0456913, the Eberly research funds of Penn State,
Polish Committee for Scientific Research (KBN) grant \mbox{2 P03B 130
  24} and the Polish Ministry of Science and Education grant 
\mbox{1 P03B 075 29}.


\appendix

\section{Electrovac NEHs in 4D spacetime}
  \label{sec:evac-geom}

The geometry of a non-expanding horizon embedded in $4$-dimensional
spacetime and admitting arbitrary matter field (satisfying certain
energy conditions) as well as the structure of its degrees of freedom
was studied in detail in \cite{abl-g}. In the analysis the
Newman-Penrose formalism occurred to be particularly convenient for the
NEH description (see especially Appendix B in \cite{abl-g}). Below we
present the geometry analysis for the NEH admitting the Maxwell field
only in order to provide necessary background for the analysis in
section \ref{sec:neh4d}. The general geometrical description
introduced in section \ref{sec:intro-geom} applies also to this class
of horizons: the constraints are given by the pullback of the
gravitational energy-momentum tensor $T_{\mu\nu}$ of the
electromagnetic field onto $\ih$
\begin{equation}\label{eq:GT}
  \Ricn{4}_{ab} - T_{ab}\ =\ 0 \ .
\end{equation}
In the case analyzed here however the above equations don't exceed
all the set of constraints. Considered set is completed by the
constraints on the electromagnetic field $F$ on $\ih$ following from
the Maxwell equations. The set of constraints extended that way is
complete: it contains all the constraints on the horizon geometry
imposed by the requirement, that $\ih$ is embedded in a spacetime
satisfying the Einstein-Maxwell field equations
\cite{Friedrich,Rendall}. 

The analysis similar to presented in this appendix was performed in
context of electrovac isolated horizons in \cite{ex}. As in there we
find particularly convenient to study the subject expressing all the 
constraints in distinguished null frame.

\subsection{The adapted frame}
  \label{sec:frame-H}

Suppose the $\ih$ is a non-expanding horizon embedded in a
$4$-dimensional spacetime and $\ell$ is a null field tangent to it and
such that its surface gravity $\sgr{\ell}$ is a constant of the
horizon. Let $v$ be a coordinate compatible with $\ell$. Then the
vector field $n^{\mu}:=-g^{\mu a}D_av$ (where $g^{\mu\nu}$ is an
inverse spacetime metric) is a null vector transversal to
the horizon and orthogonal to the constancy surfaces $\slc_v$ of $v$
(see section \ref{sec:nes-geom}).

Let $e_{\boldsymbol{\mu}} = (e_1,\,e_2,\,e_3,\,e_4) = (m,\,\bar{m},\,
n,\, \ell)$ be a
complex Newman-Penrose null frame defined in a spacetime neighborhood
of $\ih$ (see \cite{np} for the definition and basic properties). The
spacetime metric tensor and the degenerate metric tensor $q$ induced
on $\ih$ take in that frame the following form:
\begin{subequations}\label{eq:NP-g}\begin{align}
  g\ &=\ e^1\otimes e^2 + e^2\otimes e^1
      - e^3\otimes e^4 - e^4\otimes e^3 \ , \\
  q\ &=\  (e^1 \otimes e^2 + e^2\otimes e^1)_{(\ih)}.
\end{align}\end{subequations}
where $(\cdot)_{(\ih)}$ denotes the pull-back onto $\ih$.

The real vectors $\Re(m)$, $\Im(m)$ are (automatically) tangent
to $\ih$.  To adapt the frame further, we assume the vector fields
$\Re(m), \Im(m)$ are tangent to the surfaces $\slc_v$ and Lie dragged
by the flow $[\ell]$
\begin{equation}\label{m}
  \lie_{\ell} m\ =\ 0 \ .
\end{equation}
The projection of $m$ onto $\bas$ uniquely defines then on a horizon
base space a null vector frame $(\hm,\hmb)$
\begin{equation}
  \spi_* m \ =:\ \hm \ ,
\end{equation}
and the differential operator
\begin{equation}\label{eq:delta_def}
  \hmo := \hm^A\partial_A \
\end{equation}
corresponding to the frame vector $\hm$.

The frame $(e^1, e^2, e^3, e^4)$ is adapted to: the vector field $\ell$,
the $[\ell]$ invariant foliation of $\ih$, and the null complex-valued
frame $\hm$ defined on the manifold $\bas$. Spacetime frames
constructed in this way on $\ih$ will be called {\it adapted}.

Due to \eqref{m} and the normalization of $n^{\mu}$ all the
elements of an adapted frame are Lie dragged by $\ell$,
\begin{equation}
  \lie_{\bsl} e^{\boldsymbol{\mu}}_{(\ih)}\ = \ 0 \ .
\end{equation}
In consequence, the connection defined by the horizon covariant
derivative $D$ in that frame can be decomposed the following way
\begin{subequations}\label{G}\begin{align}
  \label{G12}
    m^\nu D \bar{m}_\nu\ &=\
    \spi^* \left({\hm}^A\hD {\hmb}{}_A\right)\ =: \spi^*\hat{\Gamma},\\
  \label{G43}
    -n_\nu D \ell^\nu\ &=\ \w{\bsl} =
    \bpi e^2_{(\ih)} + \bar{\bpi}e^1_{(\ih)}
    + \sgr{\bsl} e^3_{(\ih)},\\
  \label{G32}
    -\bar{m}^\nu D n_\nu\ &=\  \mu e^1_{(\ih)} +\lambda e^2_{(\ih)}
    + \bpi e^4_{(\ih)},\\
  m_\mu D \ell^\mu\ &= \ 0,
\end{align}\end{subequations}
where $\hat{\Gamma}$  is the Levi-Civita connection
1-form corresponding to the covariant derivative $\hD$
defined by $\hq$ and to the null frame $(\hm,\hmb)$ defined on $\bas$
\begin{equation}\label{2gamma}
  \hat{\Gamma}\ =:\ 2\bar{a} \hat{e}^1 + 2{a} \hat{e}^2 \ .
\end{equation}

The rotation 1-form potential $\w{\bsl}$ in the chosen
frame takes the form
\begin{equation}\label{eq:w-frame}
  \w{\bsl}\ =\  \bpi e^2_{(\ih)} + \bar{\bpi}e^1_{(\ih)}
            - \sgr{\ell}e^4_{(\ih)}\ ,
\end{equation}

In case $\bas = S^2$ the Hodge decomposition \eqref{eq:Hodge} of the
the projective rotation 1-form $\hw{\bsl}$ corresponding to $\w{\ell}$
simplifies significantly, namely one can express $\hw{\bsl}$ (and its
coefficient $\bpi$ in the decomposition \eqref{eq:w-frame}) in terms
of two real potentials $U,B$ defined on $\bas$
\begin{subequations}\label{eq:UBa}\begin{align}
  \hw{\bsl}\ &=\ \hat{\star}\hd U + \hd\ln B \ , &
  \bpi \ &=\ -i\hmob U + \hmob\ln B \ , \tag{\ref{eq:UBa}}
\end{align}\end{subequations}
where $\hat{\star}$ is Hodge star defined by the 2-metric tensor $\hq$.

The remaining two connection coefficients $(\mu,\lambda)$ (being the
only $v$ dependent ones) are the components of the transversal
expansion-shear tensor $\tS_{AB}$:
\begin{subequations}\label{eq:mulS}\begin{align}
  \mu\ &=\ \tS_{AB}\tilde{m}^A\tilde{\bar{m}}^B \ , &
  \lambda\ &=\ \tS_{AB}\tilde{\bar{m}}^A\tilde{\bar{m}}^B \
  , \tag{\ref{eq:mulS}}
\end{align}\end{subequations}
where by $\tilde{m}$ we denote the projection of $m$ onto surface
$\slc_v$.

The constraints induced by the Einstein equations \eqref{eq:GT} are by
the identity \eqref{eq:S_ev} equivalent to the following set of
equations
\begin{subequations}\label{eq:constr-evac-4D}\begin{align}
  \label{Dmu}
    T_{m\bar{m}} \ = \ \Ricn{4}_{m\bar{m}}\
    &=\ 2D\mu + 2\sgr{\bsl}\mu - \tdiv\tw{\ell}
        - |\tw{\ell}|^2_{\tq} + K \ ,\\
  \label{Dlambda}
    T_{\bar{m}\bar{m}} \ = \ \Ricn{4}_{\bar{m}\bar{m}}\
    &=\ 2D\lambda + 2\sgr{\ell}\lambda - 2\mb \bpi
        - 4a\bpi - 2\bpi^2 \ ,
\end{align}\end{subequations}
where $D:=\ell^{a}\partial_a$, $\m:=m^a\partial_a$ and
$(K,\tdiv\hw{\ell})$ are the curvature of the metric $\tq$ induced on
each surface $\slc_v$ and the divergence of projected rotation 1-form
respectively. Both the quantities are due to \eqref{eq:hq} and
the $0$th Law \eqref{eq:0th} pullbacks of the corresponding ones
$(K,\hdiv\hw{\ell})$ defined on the horizon base space.
\begin{subequations}\label{eq:Kdw}\begin{align}
  K \ &:= \ 2\hmo a + 2\hmob \bar{a} - 8 a  \bar{a} \ ,  \\
  \hdiv\hw{\ell}\
    &= \ \hmo\bpi + \hmob\bar{\bpi} - 2a\bar{\bpi} - 2\bar{a}\bpi \ .
\end{align}\end{subequations}

\subsection{Einstein-Maxwell constraints}
  \label{sec:4D-evac-constr}

The constraints on horizon geometry imposed by Maxwell equations
were analyzed in detail in \cite{ex}. In geometric form they can be
written down as the following set of equations
\begin{subequations}\label{eq:M-constr}\begin{align}
  \ell^{\mu}(\star\left(\rd\cF\right))_{\mu\nu}\ &=\ 0 \ , &
  ({}^{(\ih)}\star\ -\ i)\left[\left(\star \rd\cF\right)_{(\ih)}
    \right]\ &=\ 0 \ ,
  \tag{\ref{eq:M-constr}}
\end{align}\end{subequations}
where $\star$ and ${}^{(\ih)}\star$ are, respectively, the spacetime
Hodge star and a Hodge dual intrinsic to the horizon, and $\cF$ is the
self-dual part $\cF$ of electromagnetic field tensor $F$
\begin{equation}\label{eq:cF-def}
  \cF\ :=\ F-i\star F \ .
\end{equation}

The equation \eqref{eq:lF} and the metric decomposition
\eqref{eq:NP-g} imply that the energy momentum tensor component
$T_{\ell\ell}$ vanishes at $\ih$, thus the Maxwell field $F$ satisfies
the Weaker Energy Condition \eqref{eq:Tll>0}. Hence from the Einstein
equation \eqref{eq:GT} and the Raychaudhuri equation follows that
$T_{ab}\ell^a\ell^b = 0$. A consequence of this fact is
\begin{equation}\label{eq:lF}
  \ell\lrcorner \cF_{(\ih)}\ =\ 0.
\end{equation}
Therefore, if we consider the tensor $\cF_\nu{}^\mu$ as a 1-form
taking  vector values, then its pullback on $\ih$ takes values in the
space $T\ih$ tangent to $\ih$.

Concluding, an {\it electrovac non-expanding horizon} is a NEH $\ih$
which admits  an electromagnetic field $F$ such that the constraint
equations \eqref{eq:M-constr}, \eqref{eq:lF} are satisfied. 

Equation \eqref{eq:lF} implies that $\ell^a T_{ab}=0$, ensuring the
satisfaction of the Stronger Energy Condition \ref{c:energy}, so the
$0$th Law \eqref{eq:0th} (via the constraint equations). Also the
equations (\ref{eq:M-constr}a, \ref{eq:lF}) allow us to write the
pull-back of $\cF$ to $\ih$ as a pull-back of the field tensor $\hcF$
defined on the horizon base space
\begin{equation}\label{eq:hcF}
  \cF_{(\ih)}\ =\ \spi^*\hcF \ ,
\end{equation}
and denoted as the self-dual part of the {\it projective
  electromagnetic field} tensor.

Given an electromagnetic field $F=\frac{1}{2}F_{\mu\nu}e^\mu\wedge
e^\nu$  present in a spacetime neighborhood of the horizon in a null
frame proposed in the previous section it can be decomposed as follows,
\begin{equation}\label{F}
  F\ =\ -\Phi_0e^4\wedge e^1 + \Phi_1(e^4\wedge e^3 + e^2\wedge e^1)
  - \Phi_2 e^3\wedge e^2 + c.c. \ .
\end{equation}
The components of the Ricci tensor appearing in the
equations \eqref{eq:constr-evac-4D} are then (according to the
Einstein field equations) equal to
\begin{subequations}\label{eq:cevRic}\begin{align}
  \Ricn{4}_{mm}\ &=\ 0, &
  \Ricn{4}_{m\bar{m}}\ &=\ 2|\Phi_1|^2. \tag{\ref{eq:cevRic}}
\end{align}\end{subequations}
respectively.

Condition \eqref{eq:lF} implied by the constraints reads
\begin{equation}\label{Phi0}
  \Phi_0 \ =\ 0 \ .
\end{equation}
whereas the part of the constraints \eqref{eq:M-constr} coming from
the Maxwell equations forms the following system involving $\Phi_1,
\Phi_2$:
\begin{subequations}\begin{align}
  \label{eq:DPhi1}
    D\Phi_1\ &=\ 0 \ ,  \\
  \label{eq:DPhi2}
    D\Phi_2\ &=\ -\sgr{\bsl}\Phi_2 + (\mb+2\bpi)\Phi_1 \ .
\end{align}\end{subequations}

Due to \eqref{eq:DPhi1} the component $\Phi_1$ is a pullback of the
complex coefficient defined on $\bas$
\begin{equation}\label{eq:Phi1-pull}
  \Phi_1\ =:\ \spi^*\hat{\Phi}_1 \ .
\end{equation}
The projective field tensor defined via \eqref{eq:hcF} can be then
written as
\begin{equation}\label{eq:hcF-dec}
  \hcF\ =\ i\hat{\Phi}_1\hvol \ ,
\end{equation}
where $\hvol$ is an area form of $\bas$.

$\hat{\Phi}_1$ is invariant with respect to both: transformation of
null field $\ell\to f\ell$ and change of the coordinate compatible
with $\ell$: $v\to v + \spi^*\hat{v}_o$.

Suppose now the NEH admits an extremal null infinitesimal
symmetry $\bsl$. The construction specified in section \ref{sec:frame-H}
can be then used to define null frame such that $e_4 = \bsl$,
well defined on (dense due to theorem \ref{thm:ell-ex}) subset
$\mathcal{U}$ of $\ih$ on which $\bsl\neq 0$. Due to (\ref{Phi0},
\ref{eq:DPhi1}) at $\mathcal{U}$ the condition \eqref{eq:lielF} is
equivalent to $D\Phi_2=0$, hence the Maxwell equation \eqref{eq:DPhi2}
takes the form
\begin{equation}\label{eq:pre-ex-M-c}
  (\mb + 2\bpi)\Phi_1\ =\ 0 \ .
\end{equation}
All the extremal null infinitesimal symmetries are necessarily
of the form $\bsl = (\spi^*\hat{B})\ellJK$, where $\ellJK$ is
Jezierski-Kijowski null vector field (defined globally on $\ih$). One
can then rewrite \eqref{eq:pre-ex-M-c} in terms of coefficients in
frame (again constructed as specified in section \ref{sec:frame-H})
such that $e_4 = \ellJK$.

As at the horizon $\Phi_0 = D\Phi_1 = 0$ all the tetrad
transformations not changing the direction of $e_4$ (see for example
\cite{exact} for the classification of frame transformations as well
as corresponding coefficients transformation rules)\footnote{Without
  loose of generality we can
  exclude transformations $m\mapsto e^{i\theta}m$.} leave  both
$\Phi_1$ and $\mb\Phi_1$ unchanged. On the other hand upon
change $\ell\mapsto\ell'=(\spi^*\hat{B})\ell$ the coefficient
$\bpi$ is modified the following way
\begin{equation}
  \bpi \to \bpi' = \bpi + \mb\ln(\spi^*\hat{B}) \ .
\end{equation}
In the frame such that $e_4 = \ellJK$ the constraint
\eqref{eq:pre-ex-M-c} will take then the form:
\begin{equation}\label{eq:ex-M-c}
  [\mb + 2\bpi + 2(\mb\ln(\spi^*\hat{B})) ]\Phi_1\ =\ 0 \ .
\end{equation}
Above equation involves objects well defined on the entire $\ih$
so it holds on the closure of $\mathcal{U}$, thus globally on the
horizon.

\section{Coordinates adapted to axial symmetry on a $2$-sphere}
  \label{sec:axi-coord}

Consider a $2$-dimensional manifold $S$ diffeomorphic to the
sphere and equipped with a metric tensor $\hq_{AB}$. Suppose $S$
admits an axial symmetry group generated by the field
$\hat{\Phi}^A$. One can then introduce on that manifold a distinguished
coordinate system defined in terms of the geometric objects only. Such
a construction has been developed in \cite{ex,aepv} and has proved to
be useful in various applications (analysis of axial solutions,
construction of multipole decomposition of an IH geometry). Here we
will present th construction and analyze in detail conditions for the
global definiteness (in particular differentiability) of the metric
on the sphere. They will be formulated as conditions on the
coefficients representing $\hq_{AB}$ in considered coordinate system.

Denote the area form and radius of $S$ by $\hvol$ and $R$
respectively (where $R$ is defined via manifold area $A=4\pi R^2$).
Given $\hvol$ and the axial symmetry field $\hat{\Phi}^A$ there exists
a function $x$ globally defined on $S$ and such that\footnote{We
  follow the convention of \cite{aepv}.}
\begin{subequations}\label{eq:x-def}\begin{align}
  \hD_A x\ &:=\ \frac{1}{R^2}\hvol_{AB}\hat{\Phi}^A \ , &
  \int_S x\hvol\ &=\ 0 \ .
\end{align}\end{subequations}
By the definition $\lie_{\hat{\Phi}}x=0$ and $\hD_Ax$ vanishes only at
the poles. Hence $x:\ S\mapsto [-1,1]$ is a function monotonically
increasing from one pole to another.

Let us now introduce on $S$ the vector field $x^A$ such that
\begin{subequations}\begin{align}
  \hq_{AB} x^A \hat{\phi}^A\ &=\ 0 \ ,  &
  x^A\hD_a x\ &=\ 1 \ .
\end{align}\end{subequations}
Such a field (well defined everywhere except the poles) necessarily
takes the form
\begin{equation}
  x^A\ =\ \frac{R^4}{|\hat{\Phi}|^2}\hq^{AB}\hD_A x \ .
\end{equation}
Given the field $x^A$ one can define the coordinate $\varphi$
compatible with $\hat{\Phi}^A$ the
following way:
\begin{itemize}
  \litem Choose on $S$ a single integral curve $\hat{\gamma}$ of $x^A$
    connecting the poles. Set the function $\varphi$ to $0$ on
    $\hat{\gamma}$.
  \litem As the field $x^A$ was defined in terms of the geometric
    objects only an action of an axial symmetry maps one integrate
    curve of $x^A$ onto another. We can then extend the coordinate
    $\varphi$ attaching to each point of $S$ (except the poles) the
    value of group parameter needed for mapping of $\hat{\gamma}$ into
    curve intersecting given point.
\end{itemize}
The pair $(x,\varphi)$ will be referred to as the coordinate system
adopted to an axial symmetry.

The metric tensor in the coordinate system defined above is the
following
\begin{subequations}\label{eq:hq-Coord}\begin{align}
  \hq_{AB}\ &=\ R^2(P^{-2}\hD_Ax\hD_Bx + P^2\hD_A\varphi\hD_B\varphi)
  &
  \hq^{AB}\ &=\ \frac{1}{R^2}(P^2x^Ax^B
    + P^{-2}\hat{\Phi}^A\hat{\Phi}^B)
    \tag{\ref{eq:hq-Coord}}
\end{align}\end{subequations}
whereas the $2$-dimensional Ricci tensor takes the very simple form
\begin{equation}\label{eq:axi-Ric}
  \Ricn{2}(x,\varphi)\ =\ - \frac{1}{R^2}\partial_{xx}P(x)^2 \ .
\end{equation}
The function $P:=\fracs{1}{R}|\hat{\Phi}|$ will be referred to as the
{\it frame coefficient}.

The coordinates defined above are not well-defined at the poles, thus
the formulation of a smoothness condition for the metric at those points
requires careful analysis as the norm of $\hat{\Phi}$ (so $P$)
vanishes there. Also $\varphi$ has a $2\pi$ discontinuity on one
integral curve $\hat{\gamma}$ of $x^A$ which is however a standard
discontinuity of an angle coordinate thus is not problematic.

On the whole sphere except the poles the necessary and sufficient
condition for smoothness and well-definiteness of the metric is the
smoothness and explicit positiveness of $P$. On the other hand the
requirement of absence of conical singularities at the poles
imposes non-trivial condition for $P$
\begin{equation}\label{eq:axi-glob}
  \lim_{x\to \pm 1}\, \partial_x P^2\ =\ \mp 2 \ ,
\end{equation}
which uniquely determines $P$ for given $\Ricn{2}$
\begin{equation}\label{eq:P-Ric}
  P^2\ =\ 2(x+1) - R^2\int^{x}_{-1}\int^{x'}_{-1}\Ricn{2}\hd x''\hd x'
    \ .
\end{equation}
In fact the conditions: \eqref{eq:axi-glob}, $P|_{\pm1}=0$ and
requirement of smoothness of $\Ricn{2}$ are sufficient for smoothness
of $\hq$ at the poles. Indeed the following is true:
\begin{theorem}\label{thm:axi-glob}
  Suppose $P:[-1,1]\mapsto\re$ such that $P^2\in C^k([-1,1])$
  satisfies the following conditions:
  \begin{subequations}\label{eq:axi-glob1}\begin{align}
    \forall_{x\in ]-1,1[} \quad P(x)\ &>\ 0 \ , &
    P(x=\pm 1)\ &=\ 0 \ , &
    \lim_{x\to \pm 1}\, \partial_x P^2\ &=\ \mp 2 \ , &
       \tag{\ref{eq:axi-glob1}}
  \end{align}\end{subequations}
  Then the tensor $\hq_{AB}$ defined via \eqref{eq:hq-Coord} is
  positively definite axi-symmetric $k$-times differentiable metric
  tensor of a sphere. The pair $(x,\varphi)$ constitutes the
  coordinate system adapted to axial symmetry.
\end{theorem}

\begin{proof}
To proof the theorem it is enough to show that the function
$\theta\in[0,\pi[$ such that $x=:\cos(\theta)$ is proper angle
coordinate on the sphere. To do so we will analyze the relation of
proposed coordinate system with the conformally spherical one, that is
the pair $(\vartheta, \varphi)$ such that
\begin{equation}\label{eq:q_sph}
  \hq_{AB}\ =\ \tilde{P}^2(\hD_A\vartheta\hD_B\vartheta
           + \sin^2(\vartheta)\hD_A\varphi\hD_B\varphi) \ .
\end{equation}
The comparison of (\ref{eq:hq-Coord}a) and \eqref{eq:q_sph} allows us
to relate $P,\tilde{P}$ and $\theta,\vartheta$:
\begin{subequations}\label{eq:con}\begin{align}
  RP\ &=\ \tilde{P} \sin(\vartheta) \ , &
  \frac{R}{P}\hD_Ax\ &=\ \tilde{P}\hD_A\vartheta \ .
    \tag{\ref{eq:con}}
\end{align}\end{subequations}
On the other hand the requirement for $q_{AB}$ to be $k$ times
differentiable is equivalent to the requirement that $\tilde{P}$ is:
finite, strictly positive and $k$ times differentiable with respect to
$\vartheta$. Let us then show that these conditions are indeed
satisfied provided assumptions of theorem \ref{thm:axi-glob} hold.

For the convenience we will express the conditions \eqref{eq:con} in
terms of an auxiliary coefficient $F$ such that
\begin{equation}\label{eq:aux}
  F\ :=\ \frac{P^2}{1-x^2}  \ .
\end{equation}
They read
\begin{subequations}\label{eq:Pcoord}\begin{align}
    \tilde{P}^2\ &=\ R^2F\frac{\sin(\theta)}{\sin(\vartheta)} \ , &
    \frac{d\theta^2}{\sin(\theta)}\
      &=\ F^2\frac{d\vartheta^2}{\sin(\vartheta)} \ .
  \tag{\ref{eq:Pcoord}}
\end{align}\end{subequations}
The positiveness of $P$ everywhere except the poles implies that $F$
is also finite and strictly positive there, whereas the necessary
conditions for smoothness at the poles (\ref{eq:axi-glob1}b,c)
determine the limit of $F$ at them
\begin{equation}\label{eq:limF}
  \lim_{x\to\pm 1} F\
  =\ \lim_{x\to\pm 1} \frac{-\partial_x P^2}{2x} = 1 \ .
\end{equation}
Thus
\begin{rem}\label{thm:F_def}
  Auxiliary frame coefficient $F$ is finite and positively
  definite on $S$. In particular $F=1$ at the poles.
\end{rem}
This result allows us to establish at least boundedness and positive
definiteness of $\tilde{P}$ provided $0<|\sin(\theta)/\sin(\vartheta)|<
\infty$. To verify this condition it will be more convenient to
introduce 'plane equivalents' $\tilde{t},t$ of coordinates $\vartheta,
\theta$:
\begin{subequations}\label{eq:plane}\begin{align}
  \tilde{t}\ &=\ \ln(\tan(\frac{\vartheta}{2})) \ , &
  t\ &=\ \ln(\tan(\frac{\theta}{2})) \ .
    \tag{\ref{eq:plane}}
\end{align}\end{subequations}
An integration of (\ref{eq:Pcoord}b) gives us the relation between
$t,\tilde{t}$
\begin{equation}\label{eq:ttt}
  \tilde{t}(t)\ =\ \int_{0}^{t}\frac{dt'}{F(t')} + \tilde{t}_0 \ .
\end{equation}
As (due to \eqref{eq:limF}) $\lim_{t\to\pm\infty} F = 1$ for
sufficiently large $|t'|$ the integrand $1/F(t')$ in \eqref{eq:ttt} is
bounded from below by some positive value. Thus $\lim_{t\to\pm\infty}
\tilde{t}(t) = \pm\infty$ so one can express $\sin(\theta) /
\sin(\vartheta)$ at the poles by $\frac{d\theta}{d\vartheta}$ which is
equal to
\begin{equation}
  \frac{d\theta}{d\vartheta}\
  =\ \frac{\theta_{t}dt}{\vartheta_{\tilde{t}}d\tilde{t}}\
  =\ F\frac{\cosh(\tilde{t})}{\cosh(t)} \ ,
\end{equation}
and is finite and strictly positive on $S$ except the poles
according to Remark \ref{thm:F_def} and finiteness of $t$.
On the other hand due to explicit positiveness of $F$ $\tilde{t}$ is
finite whenever $t$ is. That implies via (\ref{eq:plane},
\ref{eq:Pcoord}a) the positivity and finiteness of $\tilde{P}$ outside
poles.

The value of $\frac{d\theta}{d\vartheta}$ at the poles
is given by the following limit:
\begin{equation}
  \lim_{t\to\pm\infty} \frac{d\theta}{d\vartheta}\
  =\ \exp( \lim_{t\to\pm\infty} |\tilde{t}(t)-t| ) \ ,
\end{equation}
where
\begin{equation}\label{eq:t_int}
  \lim_{t\to\pm\infty} ( \tilde{t}(t)-t )\
  =\ \int_0^{\pm\infty}\frac{F(t')-1}{F(t')}dt'\
  =\ \int_0^{\pm 1}\frac{F(x)-1}{F(x)(1-x^2)} \rd x \ ,
\end{equation}
At the poles the integrated expression takes the following values:
\begin{equation}\begin{split}
  \lim_{x\to\pm 1} \frac{F(x)-1}{F(x)(1-x^2)}\
  &=\ \mp\frac{1}{2} \lim_{x\to\pm 1} F_{x}\
    =\ \mp\frac{1}{2} \lim_{x\to\pm 1}
      \partial_x\frac{P^2}{1-x^2} \\
  &=\ \pm\frac{1}{2} \lim_{x\to\pm 1}
    \frac{ \partial_{xx}P^2 + 2F }{2x}\
  =\ \mp\frac{1}{4}( \partial_{xx}P^2|_{x=\pm 1} + 2 ) \ ,
\end{split}\end{equation}
so the integrate \eqref{eq:t_int} is finite. The following is then
true
\begin{rem}\label{thm:theta_def}
  If $P^2$ satisfying \eqref{eq:axi-glob1} is at least $2$-times
  differentiable in $x$ then the derivative
  $\frac{d\theta}{d\vartheta}$ is strictly positive and finite on
  $S$.
\end{rem}
Finally from (\ref{eq:Pcoord}a) and remark \ref{thm:F_def} immediately
follows, that the factor $\tilde{P}$ is also strictly positive and
finite at the poles (so the entire $S$).

To prove the differentiability of $\hq_{AB}$ (up to $k$th order)
it is enough to show that $F$ is $k$-times differentiable in $\theta$
and $\theta(\vartheta)$ is $k+1$ times differentiable in $\vartheta$.

Due to remark \ref{thm:theta_def} $\theta(\vartheta)$ is at least
differentiable. Its $1$st and $2$nd order derivative can be (via
\eqref{eq:Pcoord}) expressed in terms of $\partial_{\vartheta}\theta$
and auxiliary coefficient $F$ (and its derivative)
\begin{subequations}\label{eq:1st2nd}\begin{align}
  \frac{\hd\theta}{\hd\vartheta}\ &=\
    F\frac{\sin\theta}{\sin{\vartheta}} &
  \frac{\hd^2\theta}{\hd\vartheta^2}\ &=\ \frac{1}{F}
    \left( \frac{\hd\theta}{\hd\vartheta} \right)^2
    \left( F_{\theta} + \frac{F-1}{\sin\theta} \right)
    \tag{\ref{eq:1st2nd}}\ .
\end{align}\end{subequations}
An action of $\partial_{\vartheta}^j$ on (\ref{eq:1st2nd}b) produces
a recursive expression for $n+2$th derivative of $\theta(\vartheta)$
which involves $\sin(\theta), \cos(\theta), F(\theta)$ and the derivatives
of $F$ over $\theta$ up to $j+1$ order (where $1$st order derivatives
over $\vartheta$ of any component were rewritten as derivatives over
$\theta$ via (\ref{eq:1st2nd}a)). As $P^2\in C^{k}([-1,1])$ the
derivatives up to $k+1$ order are continuous everywhere except the
poles. Thus to show the global differentiability one needs only to
check whether derivatives are well defined (and finite) at the
poles. The sufficient condition for that is the differentiability in
$\theta$ (up to $k$th order) of the function $F$ and term
$\fracs{F-1}{\sin\theta}$. To examine this property we will apply the
following Lemma (which for the reader convenience will be proved
later)
\begin{lem}\label{lem:int-int}
  Suppose $f:[0,a[\to\re$ is $k$-times differentiable in its domain of
  dependence and $f$ itself as well as the derivatives are finite at
  $0$. Then the following function:
  \begin{equation}\label{eq:iix2}
    \bar{\bar{f(x)}}\
    :=\ \frac{1}{x^2}\int^x_0\rd x'\int^{x'}_0 f(x'')\rd x''
  \end{equation}
  is also $k$ times differentiable at $[0,a[$ (in particular there
  exist one-sided derivatives of $\bar{\bar{f}}(x)$ at $x=0$) and
  (together with its derivatives) finite at $0$.
\end{lem}
Let us prove the differentiability of $F$ at $x=-1$ first.
By substitution of $P^2$ in \eqref{eq:aux} by \eqref{eq:P-Ric} one can
express $F$ as:
\begin{equation}
  F\ =\ \frac{1}{1-x} \left[ 2
     - \frac{R^2}{x+1} \int^x_{-1}\hd
       x'\int^{x'}_{-1}\Ricn{2}\hd x''
     \right] \ .
\end{equation}
Due to Lemma \ref{lem:int-int} the function
\begin{equation}\label{eq:bbRic}
  \bbRicn{2}\ :=\ \frac{1}{(x+1)^2} \int^x_{-1}\hd
                  x'\int^{x'}_{-1}\Ricn{2}\hd x''
\end{equation}
is $k$-times differentiable in $x$ at $x=-1$, so is $F$ as it can be
expressed as follows
\begin{equation}
  F\ =\ \frac{1}{1-x} \left[ 2
     - R^2 (x+1) \bbRicn{2}
     \right] \ .
\end{equation}
This implies that an action of the operator
$\partial_{\theta}=-(1-x^2)^{\frac{1}{2}}\partial_x$ (up to $k$ times)
produces expressions continuous at $x=-1$. One could worry that
terms $(1+x)^{\frac{1}{2}}$ produced by an action of
$\partial_{\theta}$ may produce singularity there (when differentiated) 
but it is easy to show, that they always combine with positive powers
of $(1+x)$ thus the combined terms are always of the form 
$(1+x)^{\frac{n}{2}}$, where $n$ is non-negative.

The differentiability of $F$ at $x=1$ can be shown analogously. The
only modification to the algorithm used above we need to implement is
to change the starting point of integration in \eqref{eq:P-Ric} (with
appropriate change of the remaining terms in the expression).

The term $\fracs{F-1}{\sin\theta}$ can be expressed analogously to
$F$
\begin{equation}\begin{split}
  \frac{F-1}{\sin\theta} \
  &=\ \frac{1}{(1-x)^{\frac{3}{2}}} \left[ (1+x)^{\frac{1}{2}}
     - \frac{R^2}{(x+1)^{\frac{3}{2}}} \int^x_{-1}\hd
       x'\int^{x'}_{-1}\Ricn{2}\hd x''
     \right] \\
  &=\ \frac{1}{(1-x)^{\frac{3}{2}}} \left[ (1+x)^{\frac{1}{2}}
     - R^2 (x+1)^{\frac{1}{2}} \bbRicn{2}
     \right] \ ,
\end{split}\end{equation}
hence repeating all the steps of proof of the differentiability
of $F$ we also show the differentiability at the poles (up to $k$th
order) of this term.
\end{proof}

\begin{proof}[Proof of Lemma \ref{lem:int-int}]
We need only to check the differentiability at $x=0$.
The $i$th derivative of the function $\bar{\bar{f}}$ defined via
\eqref{eq:iix2} is of the form
\begin{equation}
  \partial^i_x\bar{\bar{f}}(x)\ =\ 2(-1)^i\frac{A_i}{(i+2)!x^{i+2}} \ ,
\end{equation}
where
\begin{subequations}\label{eq:tmp-rec1}\begin{align}
  A_0\ &=\ \int^x_0\rd x'\int^{x'}_0 f(x'')\rd x''\ ,  &
  A_{i+1}\ &=\ x\partial_xA_i - (n+2)A_i \ .
  \tag{\ref{eq:tmp-rec1}}
\end{align}\end{subequations}
The second derivative of $A_i$ is always of the form
\begin{equation}\label{eq:tmp-rec2}
  \partial^{2}_x A_i = x^i \partial^{i+2}_x A_0\ .
\end{equation}
Indeed this is true for $i=0$. Moreover differentiating
\eqref{eq:tmp-rec1} twice one can show that provided
\eqref{eq:tmp-rec2} holds for $i$ it is also satisfied for
$i+1$. Therefore by induction \eqref{eq:tmp-rec2} is true for every
non-negative integer $i$.

The term $A_i$ and $\partial_xA_i$ always vanish at $x=0$. Hence by
application of del'Hospital rule twice we get
\begin{equation}
  \lim_{x\to 0} \partial^i_x\bar{\bar{f}}(x)\
  =\ \lim_{x\to 0} 2(-1)^i\frac{A_i}{(i+2)!x^{i+2}}\
  =\ \lim_{x\to 0} \frac{2(-1)^i}{(i+2)!}\partial^{i+2}_x A_0 \ .
\end{equation}
The right-hand side is finite according to the differentiability of
$f$. This completes the proof.
\end{proof}

\section{Systematic development of the Hawking field}
  \label{ssec:Hawk-syst}

Below we present a systematic method of the derivation of the Hawking
field. The calculations below are performed for a helical
infinitesimal symmetry however the derivation method is general: it
can be applied to any case of an infinitesimal symmetry generating at
the (maximally extended) horizon non-compact symmetry group and such
that the infinitesimal symmetry induced by it on the horizon base space
generates compact symmetry group there.

For the need of the development we slightly change the definition of
the Hawking field (definition \ref{def:hawking}): here by a Hawking
null field we denote a null vector field $\ell_{(X)}\Gamma\in T(\ih)$
satisfying the following conditions:
\begin{itemize}
  \item $D_{\ell_{(X)}}\ell_{(X)}=\kappa\ell_{(X)}$, where a constant
    $\kappa$ is defined below,
  \item for  every intersection $p$ between a null geodesic
    generator of $\Delta$ and  an open orbit of $X$, $\ell_{(X)}\neq
    0$ and
    \begin{equation}
      v'(U_{2\pi}(p))\ =\ v'(p) + 2\pi \ ,
    \end{equation}
    where $v'$ is a parametrization of the geodesic curve compatible
    with the vector field $\ell_{(X)}$,
  \item $\ell_{(X)}=0$ at every closed orbit of $X$.
\end{itemize}
The conditions above determine $\ell_{(X)}$ uniquely on $\ih$.

The constant $\kappa$ is defined as follows: denote by $\bar{v}$ a
function, such that $\ellJK^aD_a\bar{v}=1$ where $\ellJK$ is the
Jezierski-Kijowski vector field. Then, there is a constant
$\kappa$, and a function $b$ defined on $\ih$ such
that\footnote{$\kappa$ is related to $\sgr{X}$ via equality:
  $\kappa=-\sgr{X}$}
\begin{equation}
  U_{2\pi}^*\bar{v}\ =\ e^{2\pi\kappa}\bar{v} + b \ ,
  \quad \ellJK^aD_ab=0 \ .
\end{equation}

Now, we derive the Hawking vector field for the case $\kappa\not=0$.
If it exists, it has the following form
\begin{equation}\label{eq:Hawk-form}
  \ell_{(X)}\ =\ (\kappa \bar{v}+c)\ellJK \ ,
  \quad \kappa=\const \ ,
  \quad \ell_0^aD_ac=0 \ .
\end{equation}
Integration to this equation leads us to the following
expression for the coordinate $v'$ compatible with $\ell_{(X)}$
\begin{equation}
  v'\ =\ \frac{1}{\kappa}\ln(\bar{v}+ \frac{c}{\kappa}) + v_o \ ,
\end{equation}
where $\ellJK^aD_av_o=0$.

Upon an action of $U_{2\pi}$ the coordinate $v'$ changes as follows
\begin{equation}
  U_{2\pi}(v')\ =\ 2\pi +
  \frac{1}{\kappa}\ln\left( \bar{v}
    + e^{-2\pi\kappa}(b+\frac{c}{\kappa}) \right) + v_o \ .
\end{equation}

The desired condition $v'(U_{2\pi}(p))\ =\ v'(p) + 2\pi$ determines
the function $c$ as well defined on entire $\ih$ and differentiable
as many times as the function $b$
\begin{equation}
  c\ =\ \frac{\sgr{X}b}{1-e^{2\pi\sgr{X}}} \ .
\end{equation}
Therefore, resulting formula is determined at every $p$ such that
$U_{2\pi}^*v'\neq v'$. Remarkably, it smoothly extends to the points
$U_{2\pi}^*v'=v'$ such that $\ell_{(X)}=0$ at those points.

Since the Hawking vector field is determined just by $X$ and
$(q,D)$, it is necessarily preserved by every symmetry of
$\ih$, hence
\begin{equation}
  [X,\ell_{(X)}]\ =\ 0 \ .
\end{equation}


\end{document}